%% file: embedding_comparator.tex
\documentclass[manuscript,sigconf]{acmart}

\usepackage{subcaption}

\AtBeginDocument{%
  \providecommand\BibTeX{{%
    \normalfont B\kern-0.5em{\scshape i\kern-0.25em b}\kern-0.8em\TeX}}}

\copyrightyear{2022} 
\acmYear{2022} 
\setcopyright{rightsretained} 
\acmConference[IUI '22]{27th International Conference on Intelligent User Interfaces}{March 22--25, 2022}{Helsinki, Finland}
\acmBooktitle{27th International Conference on Intelligent User Interfaces (IUI '22), March 22--25, 2022, Helsinki, Finland}\acmDOI{10.1145/3490099.3511122}
\acmISBN{978-1-4503-9144-3/22/03}

\begin{document}

\title[Embedding Comparator]{Embedding Comparator: Visualizing Differences in Global Structure and Local Neighborhoods via Small Multiples}

\author{Angie Boggust}
\authornote{Both authors contributed equally to this research.}
\affiliation{%
  \institution{MIT CSAIL}
  \city{Cambridge}
  \state{Massachusetts}
  \country{USA}
}
\email{aboggust@csail.mit.edu}

\author{Brandon Carter}
\authornotemark[1]
\affiliation{%
  \institution{MIT CSAIL}
  \city{Cambridge}
  \state{Massachusetts}
  \country{USA}
}
\email{bcarter@csail.mit.edu}

\author{Arvind Satyanarayan}
\affiliation{%
  \institution{MIT CSAIL}
  \city{Cambridge}
  \state{Massachusetts}
  \country{USA}
}
\email{arvindsatya@mit.edu}

\input{sections/00_abstract}

\ccsdesc[500]{Human-centered computing~Visualization systems and tools}
\ccsdesc[300]{Computing methodologies~Machine learning}

\keywords{machine learning, embedding spaces, visualization system, interactive, small multiples}

\begin{teaserfigure}
  \includegraphics[width=\textwidth]{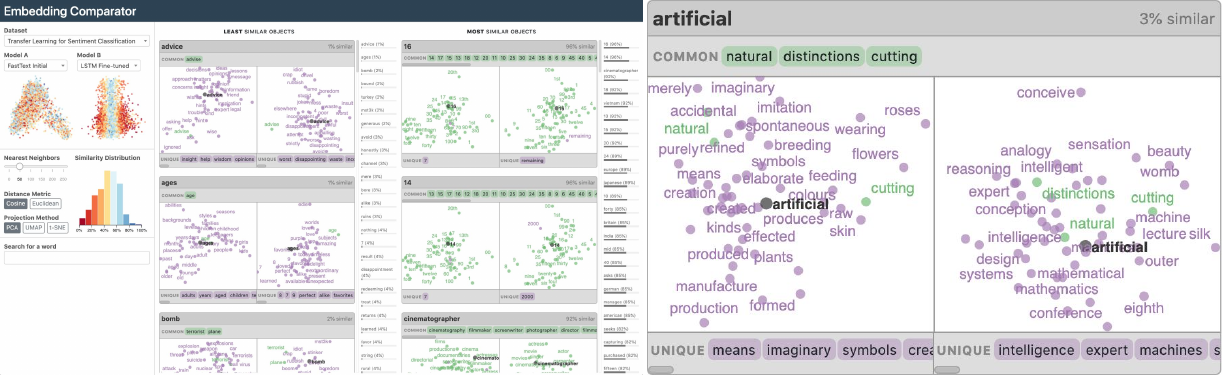}
  \caption{The Embedding Comparator (left) facilitates comparisons of embedding spaces via \emph{local neighborhood dominoes}: small multiple visualizations depicting local substructures (right).}
  \Description[The Embedding Comparator interface (left) and an example of a local neighborhood domino (right). In the left figure, the Embedding Comparator's left-hand sidebar is configured with the Transfer Learning for Sentiment Classification dataset, where Model A contains fastText embeddings, and Model B contains LSTM Fine-tuned embeddings (details of this case study in Section 5.1). The number of nearest neighbors is 50, distance metric is cosine, and projection method is PCA. No words are selected in the search bar. For additional details on visual components of the Embedding Comparator, see Figure 3. The similarity distribution is centered around 50\%, and the global projections reveal differences in global arrangement of objects as a result of fine-tuning (discussed in Section 5.1).  The local neighborhood dominoes reveal least similar objects including "advice", "ages", and "bomb" and most similar objects including "16", "14", and "cinematographer". In the right figure, an example local neighborhood domino is shown for the word "artificial". Details of local neighborhood domino visualizations are provided in Section 4.3. The common words in this domino are "natural", "distinctions", and "cutting", while unique words in model A include "means", "imaginary", and "symbols", and unique words in model B include "intelligence", "expert", and "machines". Neighborhood plots within the domino the local neighbors of "artificial" in each model and use color to encode whether each neighbor is common to both models (green) or unique to one model (purple).]{}
  \label{fig:teaser}
\end{teaserfigure}

\maketitle

\input{sections/01_intro}

\input{sections/02_related_work}
\input{sections/03_formative_interviews}
\input{sections/04_design}
\input{sections/05_case_studies}
\input{sections/06_evaluative_interviews}
\input{sections/07_conclusion}

\begin{acks}
This work is supported by a grant from the MIT-IBM Watson AI Lab.
Research was also sponsored by the United States Air Force Research Laboratory and the United States Air Force Artificial Intelligence Accelerator and was accomplished under Cooperative Agreement Number FA8750-19-2-1000. The views and conclusions contained in this document are those of the authors and should not be interpreted as representing the official policies, either expressed or implied, of the United States Air Force or the U.S. Government. The U.S. Government is authorized to reproduce and distribute reprints for Government purposes notwithstanding any copyright notation herein.
We also thank Jonas Mueller, Hendrik Strobelt, and Alan Lundgard for helpful feedback.
\end{acks}

\bibliographystyle{ACM-Reference-Format}
\bibliography{refs}

\clearpage

\appendix

\renewcommand{\thefigure}{A.\arabic{figure}}
\setcounter{figure}{0}

\input{sections/08_supplement}

\end{document}

%% file: sections/00_abstract.tex
\begin{abstract}
Embeddings mapping high-dimensional discrete input to lower-dimensional continuous vector spaces have been widely adopted in machine learning applications as a way to capture domain semantics. Interviewing 13 embedding users across disciplines, we find comparing embeddings is a key task for deployment or downstream analysis but unfolds in a tedious fashion that poorly supports systematic exploration. In response, we present the Embedding Comparator, an interactive system that presents a global comparison of embedding spaces alongside fine-grained inspection of local neighborhoods. It systematically surfaces points of comparison by computing the similarity of the $k$-nearest neighbors of every embedded object between a pair of spaces. Through case studies across multiple modalities, we demonstrate our system rapidly reveals insights, such as semantic changes following fine-tuning, language changes over time, and differences between seemingly similar models. In evaluations with 15 participants, we find our system accelerates comparisons by shifting from laborious manual specification to browsing and manipulating visualizations.
\end{abstract}

%% file: sections/01_intro.tex
\section{Introduction}
\label{sec:intro}

Embedding models map high-dimensional discrete objects into lower-dimensional continuous vector spaces such that vectors of related objects are located close together.
Although the individual dimensions and structure of embedding spaces can be challenging to interpret, embeddings have become widely used in machine learning (ML) applications because their structure usefully captures domain-specific semantics.
For example, in natural language processing (NLP), embeddings map words into real-valued vectors in a way that co-locates semantically similar words~\cite{mikolov2013distributed}.

A critical task when working with embedding models is evaluating their learned representations.
For instance, users may wish to determine if embeddings can transfer to a low-resource domain (e.g., applying general English embeddings to medical text~\cite{howard2018universal}).
In speech recognition~\cite{bengio2014word}, computer vision~\cite{netdissect2017}, recommendation~\cite{koren2009matrix}, computational biology~\cite{bepler2018learning, bileschi2019using}, multimodal learning~\cite{rouditchenko2020avlnet}, and computational art~\cite{engel2017neural, ha2018a}, evaluating embeddings has informed training procedures and revealed the impact of different training datasets, model architectures, hyperparameters, and model initializations.

To understand how people evaluate and compare embeddings, we conducted a series of semi-structured interviews with users across disciplines who frequently use embedding models as part of their research or in application domains (Section~\ref{sec:formative-interviews}).
Users balance between examining global semantic structure via dimensionality reduction plots and inspecting local neighborhoods of specific embedded objects.
Our conversations reveal shortcomings of these approaches, including unprincipled object selection strategies that rely heavily on domain knowledge or repetitive ad hoc analysis and siloed tools that focus on either one model at a time or depict only the global structure of the embedding space.
As a result, users feel concerned that they may miss unexpected insights or lack a comprehensive understanding of the embedding space.
Moreover, users cannot develop tight feedback loops or rapidly iterate between generating and answering hypotheses as their current processes include limited interactive capabilities and, thus, require tedious manual specification.

In response, we present the \emph{Embedding Comparator}, an interactive system for analyzing a pair of embedding models.
Drawing on the insights from our formative interviews, the Embedding Comparator balances between visualizing the embeddings' global structures with comparing the local neighborhoods  (Section~\ref{sec:design}). 
To easily identify the similarities and differences between the two models, the system calculates a similarity score for every embedded object based on its reciprocal local neighborhood (i.e., the number of nearest neighbors an object shares between the two models and how many are unique).
These scores are visualized in several ways, including through a histogram of scores, through color-encoding the global geometry plots, and critically, through \emph{local neighborhood dominoes}: small multiple visualizations that facilitate rapid comparisons of local substructures.
A variety of interactive mechanics help facilitate a tight iterative loop between analyzing these global and local views\,---\,for instance, by interactively selecting points in the global plots, or by searching for specific objects, users can filter dominoes, and hovering over dominoes highlights their points in the global views to provide additional context.

We demonstrate how the Embedding Comparator helps scaffold and accelerate real-world exploration and analysis of embedding spaces through case studies and first-use studies.
Using tasks based on our formative interviews, we show how our system supports use cases, such as understanding the effects of fine-tuning, conducting linguistic analysis, and exploring multimodal embeddings.
Our system design enables the replication of previously published results using only a handful of interactions and without the need for task-specific metrics.
As we demonstrate in case studies (Section~\ref{sec:case-studies}) and validate in first-use studies (Section~\ref{sec:eval-interviews}), the Embedding Comparator shifts the process of analyzing embeddings from tedious and error-prone manual specification to browsing and manipulating a series of visualizations.

The Embedding Comparator is freely available as open-source software, with source code at: \url{https://github.com/mitvis/embedding-comparator}, and a live demo at: \url{http://vis.mit.edu/embedding-comparator}.

%% file: sections/02_related_work.tex
\section{Related Work}
\label{sec:related-work}

\subsection{Machine Learning Interpretability}
ML models are widely regarded as being ``black boxes'' as it is difficult for humans to reason about how models arrive at their decisions~\cite{lipton2018mythos}.
Numerous tools help users understand model behavior~\cite{hohman2018visual}, including visualizations for specific architectures~\cite{strobelt2018lstmvis,strobelt2019s,bau2019gandissect,liu2018nlize}.
More general techniques involve evaluating input feature importance~\cite{ribeiro2016should,sundararajan2017axiomatic,shrikumar2017learning,carter2018made}, saliency~\cite{olah2018building}, or neuron activations~\cite{kahng2017cti}.
Boxer~\cite{gleicher2020boxer} compares discrete-choice classifiers, but treats them as black boxes without considering their internals.
In contrast to these methods, our focus is on comparing the \emph{representations} learned by \emph{different} models, as internal representations may differ even while input saliency or input-output behavior remains the same.
In our formative interviews (Section~\ref{sec:formative-interviews}), we found users often compare these internal representations (e.g., to identify semantic differences between hidden layers of a particular model).

\subsection{Visual Embedding Techniques and Tools}
Interpreting the representations learned at the embedding layers of ML models is challenging as embedding spaces are generally high-dimensional and latent.
In visual analytics, a variety of techniques have been developed to visualize high-dimensional data and span multiple stages of the visualization pipeline, including data transformation, visual mapping, and view transformation~\cite{liu2017}.
To reason about embedding spaces, researchers often project the high-dimensional vectors down to two or three dimensions using techniques such as principal component analysis (PCA)~\cite{jolliffe1986principal}, t-SNE~\cite{maaten2008visualizing}, and UMAP~\cite{mcinnes2018umap}.
Visualizing these projections reveals the global geometry of these spaces as well as potential substructures such as clusters, but effectively doing so may require careful tuning of hyperparameters~\cite{wattenberg2016use}\,---\,a process that can require non-trivial ML expertise.
Prior work has demonstrated that the choice of dimensionality reduction technique can impact downstream data analysis~\cite{xia2022}. Thus, the Embedding Comparator precomputes projections using PCA, t-SNE, and UMAP, and it provides a modular system design so users can use a dimensionality reduction technique of their choice. (Section~\ref{sec:global-views}).
The system defaults to PCA, which highlights the global structure of the embedding space and is deterministic, a desire of our formative interviewees (Sections~\ref{sec:formative-interviews} and~\ref{sec:design}).
Tight integration between dimensionality reduction visualizations and interactivity has been shown to be an integral component of visual analytics pipelines~\cite{sacha2017}, and we adopt this strategy in our system design (Section~\ref{sec:design}).

By default, many projection packages generate visualizations that are static and thus do not facilitate a tight question-answering feedback loop as users need to repeatedly regenerate visualizations, slowing down the exploration process.
Recently, researchers have begun to explore interactive systems for exploring embeddings including via direct manipulation of the projection~\cite{smilkov2016embedding,pezzotti2016approximated,hollt2017cyteguide}, interactively filtering and reconfiguring visual forms~\cite{heimerl2018interactive,turkay2016designing}, and defining attribute vectors and analogies~\cite{latent-space-cartography}.
While our approach draws inspiration from these prior systems, and similarly provides facilities for exploring local neighborhoods, the Embedding Comparator primarily focuses on identifying and highlighting the \emph{similarities} and \emph{differences} between different representations of embedded objects.
To do so, we compute a similarity metric for every embedded object and use this metric to drive several interactive visualizations (Section~\ref{sec:design}).

\subsection{Methods for Comparing Embedding Spaces}
\label{sec:related-work-comparing}
To compare spaces, some techniques align embeddings through linear transformation~\cite{tan2015lexical,histwords,hamilton2016cultural,mikolov2013exploiting,chen2018visual} or alignment of neurons or the subspaces they span~\cite{li2015convergent,wang2018towards}.
In contrast, the Embedding Comparator does not align the embeddings and can be used in cases where a linear mapping between the spaces does not exist, which may occur if they have different structures~\cite{mikolov2013exploiting}.
Our system exposes the objects that are most and least similar between two vector spaces via a reciprocal local neighborhood similarity metric, and local neighborhood based metrics have been shown to usefully capture differences in embedding spaces~\cite{hamilton2016cultural}.
As desired by our interviewees, our system is deterministic and faithful to the reciprocal local neighborhoods of the models (unlike transformations that rely on linear maps or weights learned by stochastic gradient optimization).

Other techniques require users to first identify and query particular objects of interest or are restricted to particular types of data or models.
Unlike parallel embeddings~\cite{arendt2020parallel}, our system does not rely on clustering or assume embeddings form clusters corresponding to semantically meaningful concepts.
EmbeddingVis~\cite{li2018embeddingvis} compares network (graph) embeddings through relationships between node metrics and embeddings.
Liu~et~al.~\cite{liu2017visual} identify differences in how word2vec and GloVe embeddings capture syntactic relationships, but do so by visualizing semantic and syntactic analogies specifically in the context of neural word embeddings.
Similarly, Liu~et~al.~\cite{latent-space-cartography} evaluate consistency of attribute and analogy vectors across latent embedding spaces. 
In contrast, our system does not rely on analogies, which may only exist for certain classes of embedding models~\cite{ethayarajh2019towards,allen2019analogies}. 
Heimerl~et~al.~\cite{heimerl2018interactive} present visualizations of nearest neighbors and co-occurrences for word embeddings to show how the meaning of a word changes over time, but do not automatically identify such objects whose representations differ most between models.
Wang~et~al.~\cite{wang2018comparison} use cosine distance to find nearest neighbors between embeddings and then arbitrarily sample words to determine if the semantic meanings differ between the models.
Other work has demonstrated differences in representations learned by convolutional neural networks (CNNs) by deconvolution of representations of sampled images~\cite{yu2014visualizing}, alignment of hidden units with interpretable concepts~\cite{netdissect2017}, or projections of representations at different layers or epochs~\cite{rauber2016visualizing}.
The Python software package \verb|repcomp|~\cite{repcomp} quantifies the difference between two embedding spaces through local neighborhood similarity but only outputs a single global similarity value between the spaces and does not permit fine-grained inspection of object-level differences.
In contrast, the Embedding Comparator computes a similarity metric for every embedded object and initializes its view to begin with objects that are the most and least similar between the two models.
As we find in our formative interviews (Section~\ref{sec:formative-interviews}), users would benefit from tools that systematically identify objects of interest, alleviating the need to sample them in a random or biased way.
Moreover, we present these objects in an interactive graphical system that facilitates rapid exploration of differences between the embedding spaces.

Concurrent with our work, \citet{heimerl2020embcomp} developed the embedding comparison system empComp. 
embComp facilitates top-down comparison of embeddings by displaying summary visualizations of embedding space differences and employing interaction to drill down to individual objects. 
Through a survey of embedding comparison tasks and motivating examples, \citet{heimerl2020embcomp} underscore the importance of local neighborhood overlap metrics to quantify embedding space differences, validating the Embedding Comparator's use of local neighborhood similarity. 
However, unlike embComp's top-down approach, the Embedding Comparator simultaneously visualizes global views of embedding structure alongside local views of individual objects and their common and unique neighbors to enable efficient analysis of both aspects (Design Goals 2 and 3). 
By interactively linking these global and local views (Design Goal 4), the Embedding Comparator permits rapid hypothesis generation and upholds known design goals of visual ML interpretability systems~\citep{gamut}.

%% file: sections/03_formative_interviews.tex
\section{Formative Interviews}
\label{sec:formative-interviews}

\input{figures/fig_interview_summary}

To understand how embedding spaces are analyzed and compared, and to identify process pitfalls and limitations of existing tools, we conducted a series of semi-structured interviews with 13 embedding users from both academia and industry.
As listed in Fig.~\ref{fig:interview-summary}, our users work in a diverse range of domains, hold a variety of titles, and use embeddings in a wide range of projects.
We recruited embedding users from within our professional network (7/13) and through an open call in our organizations and on Twitter (6/13).
Since our goal was to learn how users use embedding spaces, we required users to have experience working with high-dimensional data.
The resulting participants all worked with learned embeddings; however, many of them were not machine learning experts and used embeddings for downstream tasks in other domains.
We initially interviewed four users, and then conducted the remaining interviews throughout the development process to iteratively refine the Embedding Comparator and increase diversity of the participant pool.

We conducted semi-structured interviews with one interviewee at a time.
Each interview lasted 30--60 minutes.
We initially conducted interviews with users in person but switched to video chat due to the COVID-19 pandemic.
In each interview, we began by describing our study objective: to understand the landscape of embedding use.
We then asked users open-ended questions and encouraged users to describe their specific experience with embeddings such as \emph{``describe how you used embedding spaces in a recent project.''}
Our questions aimed to answer the following questions:
\begin{itemize}
    \item What types of projects are embeddings used in?
    \item How do embeddings help users achieve their goals?
    \item When do users compare embedding spaces?
    \item What tools and strategies are used to compare embeddings?
    \item What challenges do users face when comparing embeddings?
\end{itemize}

\subsection{User Tasks and Goals}
Through our interviews, we find users compare embeddings in consistent ways, but their tasks and goals for comparing embeddings differs significantly (Fig.~\ref{fig:interview-summary}).
Specifically, we find users fall into two categories: \emph{model-driven} users who compare embedding spaces as a way to understand model performance and behavior; and \emph{data-driven} users who study embedded representations to uncover properties of the underlying data.

\paragraph*{Model-driven users.}
Model-driven users analyze and compare embeddings as a way to develop a deep knowledge of how their models work and why.
These users are interested in questions such as \emph{what is responsible for the performance improvement between two models?} (U1--U5),
\emph{what embeddings should I use for initialization?} (U1, U4, U6), \emph{what model will perform the best on my downstream task?} (U1, U2, U4, U5, U8), and \emph{how do differences in training data affect model behavior?} (U3, U7).
To answer these questions, model-driven users perform embedding comparison tasks such as: comparing learned embeddings from different layers of the same model, comparing embeddings from different models trained on the same data, comparing learned embeddings to the ground truth, and comparing generic embeddings to those that have been fine-tuned for a particular task.

\paragraph*{Data-driven users.}
Data-driven users utilize embeddings as a way to represent and understand relationships in the data they study.
These users typically work on projects in applied machine learning domains like computational biology or historical linguistics and use embedding spaces \emph{``to investigate questions that have been made through non-computational methods''} (U13).
Example questions from data-driven users we interviewed include: \emph{what are the structural relationships between protein sequences?} (U12), \emph{what is the relationship between x-ray images and corresponding radiology reports?} (U9), and \emph{how has the plot speed in fiction novels shifted over time?} (U10).
Use cases include selecting embeddings for downstream tasks (e.g., comparing pre-trained word embeddings to embeddings fine-tuned on task-specific literature for document classification [U11]) and using embeddings to analyze differences in the data (e.g., comparing word embeddings trained on literature from different centuries [U13] or comparing embedded protein relationships to ground truth data [U12]).

\subsection{Current Processes and Challenges}
Through our interviews, we find model-driven and data-driven users use similar tools, apply similar strategies, and run into common problems when comparing embedding spaces.
A common workflow, referred to as \emph{``global check and local check''} by U5 and their colleagues, involves comparing the global structures of embedding spaces via dimensionality reduction and then selecting objects within the space and comparing their $k$-nearest neighbors.
Analyzing the global structure of embedding spaces enables users to gain insight into the semantic concepts represented by each embedding space, while comparing local neighborhoods of select objects can unearth unexpected relationships between objects or confirm user hypotheses.
For example, when studying how literary texts personify non-human characters, U11 analyzed the global structure of their embedding spaces using dimensionality reduction techniques, finding the spaces had separated into discrete clusters representing personhood identifiers: professions, education types, etc.
This finding propelled U11 to probe the local neighborhoods of known character types within each cluster like ``\textit{pilot}'' to discover new associated personifying words like ``\textit{train conductor}''.

Comparing local neighborhoods of specific objects was a critical task for the majority of our users (11/13), especially in cases when global projections were unable to meaningfully segment the space; however, many users expressed concerns that the way they selected objects to analyze was unprincipled and reduced their confidence in their results.
Data-driven users selected objects using their domain expertise (e.g., words known to change over time [U11, U13] or proteins known to interact with SARS-CoV-2 [U12]), but expressed concern that this approach would not uncover unexpected results.
Model-driven users selected objects they expected would be challenging for the model (U6), objects they hypothesized would display insightful differences (U1, U4, U5), or objects selected at random (U6); however, they were often concerned that this unsystematic approach prevented them from understanding the entire space and could be viewed as cherry-picking by the research community.
To mitigate these concerns, many users additionally compared embeddings on quantitative downstream tasks (e.g., comparing word embeddings on genre classification [U10]).
While doing so added some sense of rigor to their process, users found this procedure to be \emph{``embarrassingly empirical''} (U10) because it did not provide the same rich insights users got from local neighborhood exploration. 
As U1 emphasized, \emph{``Qualitative [analysis] is often more powerful than just a single quantitative number.''}

Throughout our interviews, users expressed concerns about the unreliability of common analysis techniques such as t-SNE and embedding alignment algorithms (U4, U7, U9, U13).
We found users distrust t-SNE due to its sensitivity to hyperparameters and stochasticity, which can lead to wildly different projections and often times misleading results~\cite{wattenberg2016use}.
Comparing embeddings using t-SNE caused users to distrust their conclusions as they were unable to draw a distinction between ``real'' findings and t-SNE artifacts.
This led U4 to stop using t-SNE altogether, noting, \emph{``You can tease apart whatever you want from t-SNE. Sometimes it shows you what you want and if it doesn't then you can spend a bit of time until you see what you want to see.''}
Users expressed similar concerns about embedding space alignment algorithms, such as orthogonal Procrustes, because the algorithms can produce unreliable and meaningless alignments (see Section~\ref{sec:related-work-comparing}).
While U13 previously applied Procrustes alignment, they are hesitant to use it again in the future because they could \emph{``only find the story [they] wanted to tell sometimes''} and struggled to determine whether their results were representative of the embedding spaces or resulted from the unpredictability of the method.

\subsection{Embedding Comparator Design Goals}
\label{sec:methods-goals}
To inform the design of the Embedding Comparator, we distill our formative interviews into the following design goals:

\renewcommand{\labelenumi}{\arabic{enumi}.}
\begin{enumerate}
    \item \textbf{Surface similarities and differences for systematic comparison.} Across all users, the central goal of comparing embedding models is to understand the degree to which they are similar, and where the differences lie.  A recurrent breakdown is how unprincipled and time-consuming identifying this information is with current approaches (e.g., performing a dimensionality reduction that does not result in informative separation or manually generating objects to analyze via $k$-nearest neighbors). Moreover, our interviews suggest that users are most interested in seeing the objects that are most similar or different\,---\,as U2 noted, \emph{``It is most interesting what happens at the extremes.''}
    
    \item \textbf{Display projection plots to surface global semantic differences between embedding spaces.} The majority of our users (9/13) use dimensionality-reduced projection plots as a primary mechanism for evaluating embedding spaces. Besides users' familiarity with these views, our interviews highlighted that visualizing global structure provides necessary context to meet our first design goal as the overall shape, density, and clustering of a projection can reveal similarities and differences in a glance. For example, U1 described how viewing the global projection of a particular embedding model caused them to stumble upon an unexpected structure they later wrote a paper about.
    
    \item \textbf{Compare local neighborhoods to identify object-level similarities and differences.} While projection plots provide useful global context, the substance of our users' analysis occurs by drilling down into and comparing the local neighborhoods of embedded objects. For example, U2 noted that even in their papers they \emph{``often report a few nearest neighbors of their models to show that they capture [meaningful] properties''}.  The Embedding Comparator must provide interface elements that facilitate rapid comparisons of reciprocal local neighborhoods for each embedded object.
    
    \item \textbf{Interactively link global and local views for rapid analysis.} Users typically explained their global and local analysis as a single technique; however, current tools force users to complete each task independently by first projecting the high dimensional space and independently querying for objects of interest. The lack of interactivity slows down their analysis processes and makes it difficult to deeply explore and identify differences that are not obvious. Thus, the Embedding Comparator should use interactive techniques to link global and local views together, including allowing users to select local views from global views, highlight local neighborhoods within the global projections, and perform targeted searches to surface the local neighborhoods of specific embedded objects.
    
    \item \textbf{Yield deterministic and reproducible results that inspire user confidence.} Our formative interviews revealed many users avoided unreliable analysis techniques such as t-SNE and Procrustes alignment. Users want to be confident the patterns they uncover are representative of true patterns in the embedding space and are not artifacts of non-deterministic tools. As such, the Embedding Comparator should leverage techniques that produce reproducible results users can trust.
\end{enumerate}

%% file: figures/fig_interview_summary.tex
\begin{figure}[t]
    \centering
    \includegraphics[width=\linewidth]{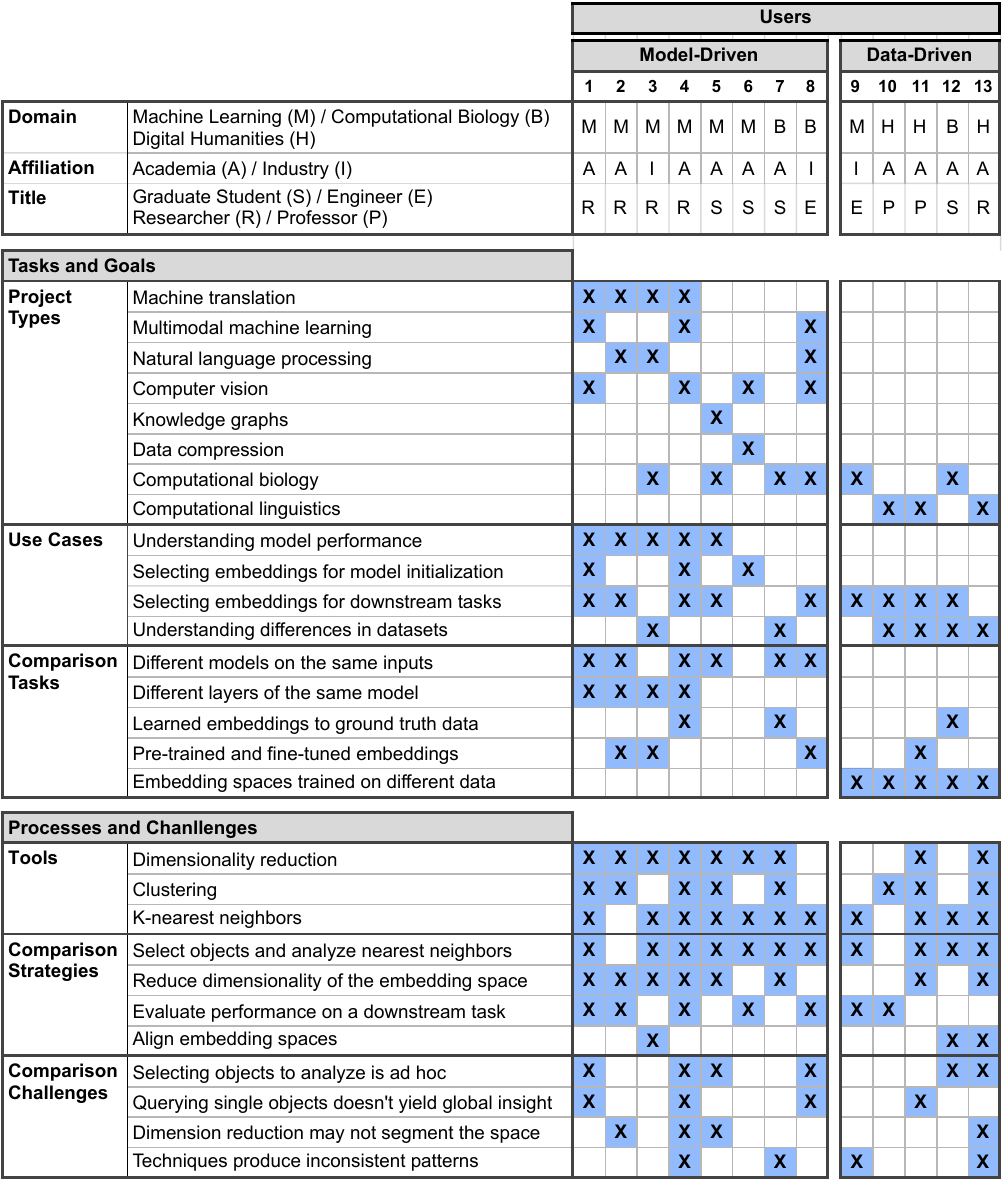}
    \caption{A matrix summary of our formative interviews grouped by user class. Users compare embedding spaces for a variety of tasks and goals, but find current processes unstructured and inconsistent due to their reliance on ad hoc exploration.}
    \Description[A matrix summary of the formative interview results. The columns are users (U1-U13). U1-U8 are model-driven users and U9-U13 are data-driven users. Row 1 describes the users' domain, affiliation, and title. U1-U6 and U9 are in machine learning. U7, U8, and U12 are in computational biology. U10, U11, and U13 are in the digital humanities. Row 2 describes their affiliation. U1, U2, U4-U7, and U10-U13 are in academia. The others are in industry. Row 3 describes the users' title. U1-U4 and U13 are researchers. U5-U7 and U12 are graduate students. U8 and U9 are engineers. U10 and U11 are professors. Rows 4-20 describe users' tasks and goals and are broken up into project types, use cases, and comparison tasks. Rows 4-11 describe project types. U1-U4 work on machine translation. U1, U4, and U8 work on multimodal machine learning. U2, U3, and U8 work on natural language processing. U1, U4, U6, and U8 work on computer vision. U5 works on knowledge graphs. U6 works on data compression. U3, U5, U7-U9, and U12 work on computational biology. U10, U11, and U13 work on computational linguistics. Rows 12-15 describe use cases. U1-U5 use embeddings to understand model performance. U1, U4, and U6 select embeddings for model initialization. U1, U2, U4, U5, and U8-U12 select embeddings for downstream tasks. U3, U7, and U10-U13 use embeddings to understand differences in datasets. Rows 16-20 describe comparison tasks. U1, U2, U4, U5, U7, and U8 compare difference models on the same inputs. U1-U4 compare different layers of the same model. U4, U7, and U12 compare learned embeddings to ground truth data. U2, U3, U8, and U11 compare pre-trained and fine-tuned embeddings. U9-U13 compare embedding spaces trained on different data. Rows 21-31 describe users' processes and challenges and are broken up into tools, comparison strategies, and comparison challenges. Rows 21-23 describe tools. U1-U7, U11, and U13 use dimensionality reduction. U1, U2, U4, U5, U7, U10, U11, and U13 use clustering. U1, U3-U9, and U11-U13 use K-nearest neighbors. Rows 24-27 describe comparison strategies. U1, U3-U9, and U11-U13 select objects and analyze the nearest neighbors. U1-U5, U7, U11, and U13 reduce dimensionality of the embedding space. U1, U2, U4, U6, and U8-U10 evaluate performance on a downstream task. U3, U12, and U13 align embedding spaces. Rows 28-31 describe comparison challenges. U1, U4, U5, U8, U12, and U13 find that selecting objects to analyze is ad hoc. U1, U4, U8 and U11 find that querying single objects doesn't yield global insight. U2, U4, U5, and U13 find that dimension reduction may not segment the space. U4, U7, U9, and U13 find techniques produce inconsistent patterns.]{}
    \label{fig:interview-summary}
\end{figure}

%% file: sections/04_design.tex
\section{System Design}
\label{sec:design}

\input{figures/fig_design_combined}

Informed by our formative interviews, the Embedding Comparator computes a similarity score for every embedded object based on its reciprocal local neighborhoods.
Critically, this similarity metric does not require the two models to have the same dimensionality nor do they need to be aligned in any way.
As a result, the Embedding Comparator is capable of supporting a wide range of embedding models.
Furthermore, the simplicity of computing overlap between reciprocal local neighborhoods makes this metric intuitive to a broad range of users, which we validate in our first-use studies (Section~\ref{sec:eval-interviews}).
To allow users to rapidly identify similarities and differences between the models, this similarity metric is visualized via a number of global views (including projection plots and a histogram distribution) as well as through \emph{local neighborhood dominoes}: small multiple views of local substructures.

\subsection{Computing Local Neighborhood Similarity}
\label{sec:methods-lns}
An embedding space is a function $E : \mathcal{V} \rightarrow \mathbb{R}^d$ mapping objects in vocabulary $\mathcal{V}$ into a $d$-dimensional real-valued vector space.
For example, in NLP, $\mathcal{V}$ may be a vocabulary of English words, and $E$ maps each word to a 200-dimensional vector.

Here, we consider two embedding spaces $E_1$ and $E_2$ over the same set of objects $\mathcal{V}$.
The embedding spaces may have different bases, a different number of dimensions $d$, or stem from different modalities.
For this reason, we compute an embedded object's similarity via the reciprocal local neighborhood around the object in each of the embedding spaces.
Precisely, for each object $w \in \mathcal{V}$, we compute the \textit{local neighborhood similarity} (LNS) of $w$ between the two embedding spaces as:
\begin{equation}
    \text{LNS}(w) = S(k\text{-NN}_1(w),\;k\text{-NN}_2(w))
    \label{eq:similarity}
\end{equation}
where $k\text{-NN}_i(w)$ returns the $k$-nearest neighbors of $w$ in embedding space $E_i$ and $S$ is a similarity metric between the two lists of nearest neighbors.
We compute $k$-nearest neighbors with cosine or Euclidean distance~\cite{turney2010frequency} and take $S$ as the Jaccard (intersection over union) similarity between the sets of neighbors.
The Jaccard similarity between two sets $C_1$ and $C_2$ is defined as $J(C_1, C_2) = \frac{|C_1 \cap C_2|}{|C_1 \cup C_2|}$ and scales between 0 (the sets are disjoint) and 1 (the sets are identical).
Users can choose between distance metrics in the interface or introduce additional functions via a JavaScript API call.
The LNS metric is deterministic (Design Goal 5), and it enables the Embedding Comparator to systematically sort and identify objects of interest, alleviating the need for users to sample them in a random or biased way or to formulate hypotheses \emph{a priori} about which objects to investigate (Design Goal 1).

The Embedding Comparator scales linearly with the number of objects. Given two $d$-dimensional embedding spaces with $|\mathcal{V}|$ objects, $k$-nearest neighbors for all objects are precomputed using an approximately $\mathcal{O}(d \cdot |\mathcal{V}| \log |\mathcal{V}|)$ algorithm (ball tree) and LNS is computed in $\mathcal{O}(|\mathcal{V}| \cdot k)$ time.
Our system is implemented in JavaScript using React and performs efficiently on the real-world case studies explored in this paper (300-dimensional embeddings and roughly 6000 objects).

\subsection{Global Views}
\label{sec:global-views}

The Embedding Comparator's left-hand sidebar (Fig.~\ref{fig:system-design}a--e) provides configuration options and interactive global views of the embedding spaces. A user begins by selecting a \emph{dataset}, which specifies the embedding vocabulary, followed by two \emph{models}, each of which defines the embedding space for that vocabulary (Fig.~\ref{fig:system-design}a).
Users can load many models, and by decoupling datasets from models, the Embedding Comparator makes it easy to compare several models with the same vocabulary.

Beneath each model, the Embedding Comparator shows a \emph{global projection} (Fig.~\ref{fig:system-design}b): a scatter plot that depicts the geometric structure of the model's embedding space, with object names shown in a tooltip on hover.
Per Design Goal 2, these projections provide valuable context during exploration and help reveal interesting global properties (e.g., distinct clusters).
Based upon our formative interviews and design goals, we load PCA projections by default because PCA is deterministic (Design Goal 5) and highlights the global structure of the embedding space.  However, we provide a modular system design, so users can interactively switch to projections computed using PCA, UMAP, or t-SNE via buttons in the sidebar.

The \emph{similarity distribution} (Fig.~\ref{fig:system-design}c) displays the distribution of LNS similarity values (Eq.~\ref{eq:similarity}) over all objects in the embedding space.
Bars are colored using a diverging red-yellow-blue color scheme to draw attention to the most extreme values (objects that are the most and least similar between the two selected models) per Design Goal 1.
We reapply this color encoding in the global projections to help users draw connections between the two visualizations and to reveal global patterns or clusters of objects with comparable similarity values (e.g., case studies of Fig.~\ref{fig:transfer-learning} and Fig.~\ref{fig:hist-words}).
Both the similarity distribution and the global projections can be used to interactively filter the dominoes (Design Goal 4; Section~\ref{sec:dominoes}), and the \emph{search bar} (Fig.~\ref{fig:system-design}e) can be used to populate the dominoes of specific object(s) of interest.

The \emph{parameter controls} (Fig.~\ref{fig:system-design}d) enable the user to interactively change the value of $k$ used to define the size of local neighborhoods for computing similarity, select the distance metric used for computing distance between vectors, and select the dimensionality reduction method used in the global projections and local neighborhood plots.
The default value of $k$ is 50, which we found to provide insightful results across our various experiments and case studies.
Changes in either of these controls immediately update the rest of the interface (Design Goal 4).

\subsection{Local Neighborhood Dominoes}
\label{sec:dominoes}

To meet Design Goal 3, the Embedding Comparator introduces \emph{local neighborhood dominoes}: a small multiples visualization to surface local substructures and facilitate rapid comparisons. 
Each domino consists of a set of interactive \textit{neighborhood plots} and \textit{common and unique neighbor lists}.

Neighborhood plots\,---\,side-by-side scatter plots that show the $k$ nearest neighbors of the domino object in each model\,---\,graphically display the relationships between the object and its neighbors.
These plots use the same projections as the global projection views (Section~\ref{sec:global-views}) to ensure all geometries are visualized consistently.
Color encodes whether each neighbor is common to both models or unique to a single model.
These neighbors are also displayed as separate scrollable lists above and below the plots and are sorted by the distance to the main object.
For example, Fig.~\ref{fig:system-design}f shows the domino for the word \emph{``mean''} from an embedding trained on text from the early 1800s (model A, left) to the 1990s (model B, right).
To facilitate cross-model comparisons, dominoes are interactive: hovering over a neighbor in the plots or lists highlights it across the entire domino (Fig.~\ref{fig:system-design}g). 
Motivated by Design Goal 1 and our users' feedback that the most interesting insights often lie at the extremes, the default view of the Embedding Comparator lists two columns of dominoes (shown in Fig.~\ref{fig:transfer-learning}): the first displays dominoes of the least similar objects (in increasing order) and the second displays those of the most similar objects (in decreasing order).
To increase information scent~\cite{pirolli2003exploring}, we adapt the Scented Widgets technique~\cite{willett2007scented} and augment the scroll bars with a domino outline (Fig.~\ref{fig:system-design}i) that shows domino objects and their similarity scores.

The dominoes' information-dense display is designed to facilitate rapid acquisition of neighborhood-level insights. 
By scanning down the dominoes, users see not only the geometries involved but also specific common and unique neighbors to trigger hypothesis generation.
Previous iterations of the Embedding Comparator used separate tabular lists
to display the most and least similar objects across models (and common and unique neighbors for individual objects) but, in early user tests, we found that such a presentation produced a high cognitive load as users tried to map back and forth between the various lists. 
Thus, with the domino design, we encapsulate all local neighborhood information associated with a given embedded object into a single interface element while still supporting cross-model comparisons.
For instance, in the \emph{``mean''} domino in Fig.~\ref{fig:system-design}f, the neighbor plots reveal substructures within the local neighborhoods\,---\,in model B, the bottom  appears to relate to the mathematical notion of \emph{``mean''}, while the top is more synonymous with \emph{``convey''} and shares neighbors with model A.
We confirm this hypothesis by scanning the common and unique lists, where we observe more mathematical words under model B than A.

\subsection{Linking Global and Local Views}

Linking the global and local views (Design Goal 4) enables users to rapidly iterate between comparing the overall embedding spaces and inspecting individual objects. 
When hovering over a domino, the object and its local neighborhoods are highlighted in each of the respective global projections with the purple/green color encoding preserved (Fig.~\ref{fig:system-design}h).
This interaction allows users to contextualize local neighborhoods within the overall embedding space.
Similarly, interactive selections (e.g., lassoing or brushing) in the global projections or similarity distribution filter the dominoes, allowing users to drill down and investigate objects of interest.

%% file: figures/fig_design_combined.tex
\begin{figure*}[t]
    \centering
    \includegraphics[width=\linewidth]{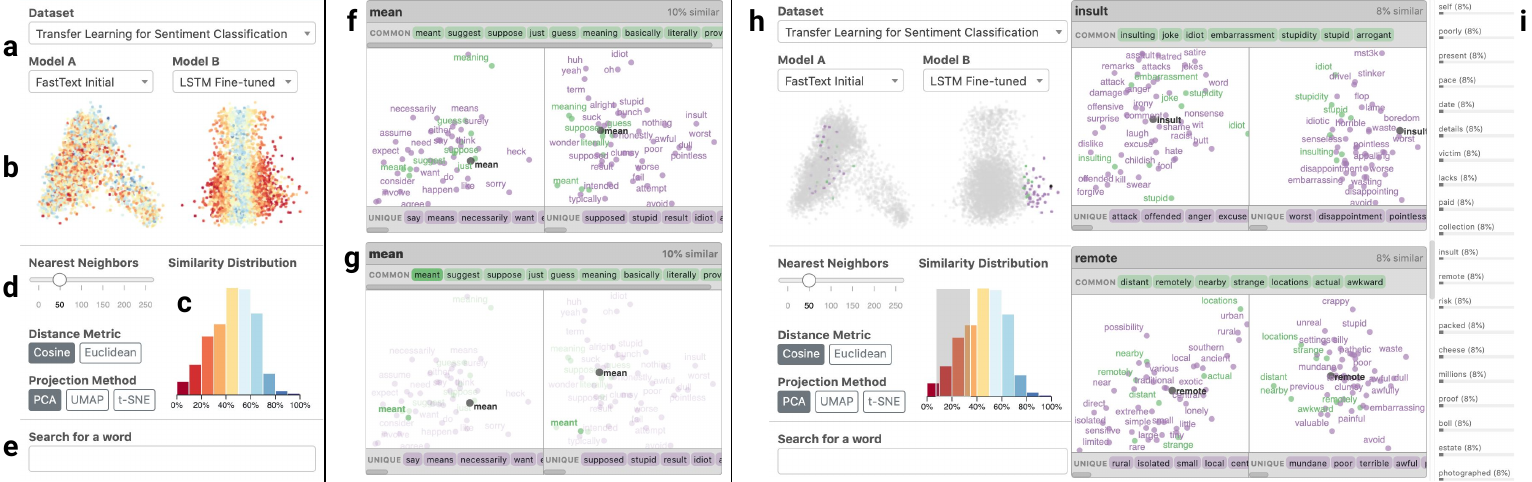}
    \caption{The Embedding Comparator interactively links global and local views. Users populate the Embedding Comparator by selecting a pair of embedding models (a). Global projections (b) visualize the geometry of the embeddings, and local neighborhood dominoes (f) visualize an object's local neighborhood. Users explore by brushing over the global projection plots (b) and similarity distribution (c), tuning the neighborhood parameters (d), searching for objects (e), and selecting dominoes from the domino outline (i). Hovering on a domino highlights its neighborhood in the global projection (h), and hovering on an object in a domino highlights that object in the local neighborhood plots (g).}
    \Description[The Embedding Comparator system diagram as described in the caption and in Section 4.]{}
    \label{fig:system-design}
\end{figure*}

%% file: sections/05_case_studies.tex
\section{Evaluation: Case Studies}
\label{sec:case-studies}

We illustrate how the Embedding Comparator accelerates real-world workflows through case studies drawn from our formative interviews (see Appendix~\ref{sec:supp-additional-case-studies} for additional examples and Appendix~\ref{sec:supp-case-studies-details} for implementation details).

\subsection{Transfer Learning for Sentiment Classification}
\label{sec:case-studies-fine-tuning}

\input{figures/fig_transfer_learning}

Our formative interviews reveal that users employ transfer learning to improve performance on downstream tasks~\cite{long2015fully,howard2018universal}, but existing tools make it difficult to compare the trade-off between pre-trained embeddings' generalizability and fine-tuned embeddings' domain-specific expressiveness.
To evaluate the Embedding Comparator on this task, we develop a transfer learning case study in collaboration with an ML researcher (model-driven user).
Here, the researcher trains an LSTM network~\cite{hochreiter1997long} to predict whether movie reviews express positive or negative sentiment.
The researcher initializes the network using pre-trained fastText English word embeddings~\cite{mikolov2018advances} and refines them using limited movie review data~\cite{imdbdataset}. 
Once complete, the researcher wishes to investigate the effect of the refinement process: for example, identifying words that are ordinarily synonyms but not in the context of sentiment prediction or words that are not ordinarily synonyms but become interchangeable in the context of sentiment prediction.

We use the Embedding Comparator to compare the pre-trained fastText word embeddings with those fine-tuned for the sentiment analysis task.
Our system immediately surfaces many insights about how the embedding space has changed as a result of fine-tuning.
The global projection plots and similarity color encoding reveal words that changed the most due to fine-tuning have moved toward the outer regions of the fine-tuned model's embedding space (Fig.~\ref{fig:transfer-learning}A).
By hovering over and zooming into the fine-tuned model's global projection, the researcher finds positive sentiment words (e.g., \textit{``favorites''}, \textit{``gem''}, \textit{``finest''}) moved toward the left side of the projection, negative sentiment words (e.g., \textit{``worst''}, \textit{``insult''}, \textit{``forgetable''}) moved toward the right, and neutral words that did not change as a result of fine-tuning (e.g., \textit{``part''}, \textit{``lights''}, \textit{``get''}) are in the center.
These interactive global views enable the researcher to identify how fine-tuning the embeddings for this classification task has affected the global arrangement of objects in the vector space: unlike the pre-trained model that disperses sentiment throughout the global space, the fine-tuned model has specifically structured the embedding space based on sentiment.

Given the significant impact fine-tuning had on the global embedding space, the researcher is now interested in how the local neighborhoods have changed.
Since inspection of the global projection plot revealed negative sentiment words on the right for the fine-tuned model, the researcher brushes the right side of the global projection plot, and the Embedding Comparator populates the dominoes with only those selected words (Fig.~\ref{fig:transfer-learning}B).
By inspecting local neighborhood dominoes, the researcher can understand how fine-tuning for sentiment analysis has affected the semantic meaning of specific objects. 
For example, in the fastText space \emph{``bore''} is most closely related to \emph{``bears''} and \emph{``carry''} but, after fine-tuning, it is most closely related to \emph{``dull''} and \emph{``waste''}, indicating fine-tuning has had the intended effect.
Next, the researcher analyzes the positive sentiment words by brushing the left side of the global projection plot.
The resultant dominoes reveal that numbers less than 10 have also been affected by fine-tuning.
For example, \emph{``7''} has become more closely related to positive adjectives as a result of numeric scales used to rank movies within the reviews.
Thus, by filtering the local neighborhood dominoes via the global projections, the Embedding Comparator exposes the types of words most affected by fine-tuning.

Intrigued by the number \textit{``7''} result, the researcher wants to understand how many other numbers have changed as a result of fine-tuning, and uses the search functionality to analyze a variety of numbers in the range 1--100 (Fig.~\ref{fig:transfer-learning}C).
Glancing at the similarity scores in the domino outline shows which numbers have changed due to fine-tuning.
Numbers typically used to convey sentiment, like 100, or those rarely used to convey sentiment, like 15, have not changed.
However, numbers that previously did not convey sentiment but do in movie review data (2--4 and 6--9) have changed significantly.
The researcher hovers over each domino to locate each number in the global projection plots and confirms the model has learned the expected sentiment for each number (Fig.~\ref{fig:transfer-learning}D).
In the fastText projection, all numbers cluster together, consistent with general language usage.
In the fine-tuned projection, 2--4 are on the negative sentiment (right) side, 6--9 are on the positive (left) side, and other numbers are in the center, consistent with the colloquial 1--10 movie review rating system.

The researcher we worked with on this case study was excited about the results the Embedding Comparator helped reveal.
Generating these types of insights with prior tools would require significant tedious effort\,---\,besides manually constructing the necessary views (e.g., within a Jupyter notebook), the researcher would have needed to formulate hypotheses \emph{a priori} about which specific words to investigate.
In contrast, by calculating a similarity score for every word, and by visualizing local neighborhoods as dominoes, the Embedding Comparator surfaces this information more directly.
As a result, the system transforms the process of comparing embedding models from requiring explicit and manual steering by the researcher, towards more of a \emph{browsing} experience. 
This shift frees the researcher to focus on generating and answering hypotheses about their models.

\subsection{Language Evolution via Diachronic Word Embeddings}
\label{sec:case-studies-histwords}

\input{figures/fig_histwords}

Our second case study follows a linguist (a data-driven user) who employs embedding models to study the evolution of languages over time.
Word embeddings can capture diachronic changes in language (i.e., changes over time)~\citep{histwords}.
Here, we use diachronic word embeddings from HistWords~\citep{histwords}, trained on English books written from 1800--2000 grouped by decade.
We select embeddings from five decades (1800--1810, 1850--1860, 1900--1910, 1950--1960, and 1990--2000) and evaluate how the Embedding Comparator surfaces words whose meanings have evolved.

Fig.~\ref{fig:hist-words}A shows the Embedding Comparator comparing embeddings of text written in 1900--1910 to text written in 1990--2000.
By scrolling through the local neighborhood dominoes, the linguist can immediately replicate known insights~\cite{histwords}, such as the change in the meaning of \emph{``gay''} (from ``happy'' to ``homosexual'') and \emph{``major''} (from ``military'' to ``important'') over the century.
The Embedding Comparator also reveals words such as \emph{``aids''}, whose meaning changed from ``assists'' to the disease HIV/AIDS, which scientists did not name until the early 1980s~\cite{centers1982update}.
In contrast to the original analysis~\cite{histwords}, with the Embedding Comparator, there is no need to align the various embedding spaces manually, nor do users need to define and compute a task-specific semantic displacement metric to uncover these findings.
Our method for comparing embedded objects through reciprocal local neighborhoods is agnostic to application domains and tasks.
As a result, the Embedding Comparator scaffolds and accelerates the analysis process for users regardless of their ML expertise.
Novice users need minimal technical knowledge to replicate established linguistic analysis, while experts can devote their effort to designing task-specific metrics only when necessary.

To further compare the two models, the linguist varies the number of neighbors $k$ using the nearest neighbors slider (Fig.~\ref{fig:hist-words}B).
Smaller values of $k$ enable comparisons of how immediate neighbors of a word have changed over time.
By dragging the slider to decrease $k$ (e.g., $k=12$), the linguist discovers words that have high similarity at small values of $k$, such as \emph{``que''}, which has a small set of French words as its nearest neighbors in both models.
This finding may be surprising to the linguist, given that the models were trained on English text.
Greater values of $k$ enable comparisons of larger neighborhoods.
Words with high similarity at larger $k$ have large neighborhoods that have remained consistent over time.
Increasing $k$ (e.g., to $k=100$) reveals words such as days and months (e.g., \emph{``april''} and \emph{``tuesday''}) whose meanings have remained consistent between the time periods.
This interaction enables the user to understand the meaning captured in the embedding spaces, particularly the size and position of different clusters.

The Embedding Comparator enables the linguist to easily compare alternate models, which is valuable in this case study since we have models trained on many decades. 
For example, using the drop-down menus, the linguist can compare 1800--1810 to 1900--1910.
In doing so, the linguist finds that during the 19th century \emph{``nice''} moves away from meaning ``refined'' and ``subtle'' and moves toward ``pleasant'', in line with previous findings~\cite{histwords}.
Similarly, by fixing one model to the most recent decade (1990--2000) and varying the other, the user can look at the similarity distribution to reveal the pairwise diachronic changes.
Immediately, the user observes that the similarity distribution is centered at 40\% and is surprised by the degree of dissimilarity of language between the 1950s and 1990s, only a 40 year period.
However, inspecting the dominoes reveals computer-related words like \textit{``artificial''} (Fig.~\ref{fig:teaser}) and \textit{``file''} have changed the most, referencing the significant technological shift over that period.
Decreasing the decade continues to shift the similarity distribution leftward, suggesting language usage diverges as time diverges (Fig.~\ref{fig:hist-words}C).
Finally, comparing the earliest decade (1800--1810) to the most recent (1990--2000) has a distribution centered at 18\% and very few words greater than 50\% similar.
Selecting the range greater than 50\% by brushing on the similarity histogram (Fig.~\ref{fig:hist-words}D) surfaces only numbers, indicating word usage changed significantly more than number usage over the past two centuries.
These interactions provide evidence for the linguist to understand how language has continuously changed over two centuries.

\subsection{Multimodal Emoji Representations}
\label{sec:case-studies-emojis}

\input{figures/fig_emojis}

Our final case study demonstrates the Embedding Comparator in a non-NLP setting: comparing emoji representations.
Since many of our formative users compared embeddings in computer vision and multimodal tasks, this case study compares emoji language embeddings (learned representations from textual emoji descriptions~\cite{eisner2016emoji2vec}) to emoji image embeddings (raw RGBA pixel values).
The model-driven user hypothesizes the language embeddings arrange emojis based on semantics, and the image embeddings arrange emojis based on visual characteristics (e.g., shape, color).
They are curious to explore if their hypothesis is correct and, if so, to understand how visual and semantic similarities are related.

The similarity distribution shown in the initial view (Fig.~\ref{fig:emojis}A) immediately reveals that emojis are represented quite differently between the two models: most emojis are 0--10\% similar, and no emojis are more than 40\% similar, suggesting visual and semantic similarity are often disjoint.
By glancing at the domino outline, the user can identify the emojis that differ the most and least between models.
While the most similar emojis are face emojis (i.e., smiling face, shocked face), the least similar emojis are more diverse objects (i.e., balloon, alligator).
This finding supports the user's hypothesis because face emojis are similar in both meaning and shape/color, whereas a least similar emoji, like balloon, does not necessarily look like its semantically similar emojis.

Digging deeper, the user looks at the baseball emoji in the least similar list (Fig.~\ref{fig:emojis}A). 
The domino for the baseball emoji shows that the nearest neighbors in the language model are other sports-related emojis (e.g., soccer ball, football), while the neighbors in the image model are emojis with similar color/shape (e.g., speech bubble, rice bowl).
The only shared neighbor between the two models is a soccer ball because it is white, round, and sports-related.
Scrolling through the sorted unique neighbors lists at the bottom of the domino lets the user see how neighbors diverge as they become more distant from the baseball.
For example, in the language model, more distant neighbors are less semantically similar\,---\,the first neighbor is a golf flag, the 13th neighbor is a dartboard, and the 50th neighbor is a gaming controller.
Similarly, in the image model, more distant neighbors are less visually similar\,---\,the first neighbor is a white circle, while the 17th neighbor is a birthday cake, and the 50th neighbor is a cow.
The domino visualization and interactions enable the user to confirm that the embedding spaces have captured their expected representations.

Given that distance between emojis corresponded to distance in meaning for the baseball emoji, the user is interested in understanding how the embedding spaces have captured meaning.
By studying other local neighborhood dominoes of the least similar objects, the user finds that the running shoe, ATM, and masked face emojis in the image model share no semantic similarity to their nearest neighbors but are similar in shape and color.
In contrast, these objects are semantically similar to their nearest neighbors in the language model.
By analyzing the masked face emoji more closely, the user finds that its nearest neighbors in the language model are related to health (e.g., hospital, apple, ambulance), and in the image model are other faces (crying face, smiling face) (Fig.~\ref{fig:emojis}B).
Hovering over the masked face domino highlights each neighborhood in the global projections.
This interaction reveals that the faces in the image embedding space form a tighter cluster than the health-related emojis in the language embedding space.
It makes sense that the faces form a tight cluster because faces are very similar images, whereas the health-related emojis are related to various other contexts (e.g., ambulances with cars, apples with food).
To confirm this finding is not an artifact of the projection method, the user switches between PCA, UMAP, and t-SNE, finding the cluster pattern persists under each technique.
Through these interactions, the user is confident that the language embeddings capture semantics learned from the emoji textual descriptions, and the image representations only capture visual similarities of the emoji images.

Next, the user shifts to scroll through the most similar objects and finds the grinning cat emoji.
The local neighborhood plots within this domino reveal two distinct clusters of emojis in the language model (happy face emojis and animal emojis) and two clusters in the image model (face emojis and cat emojis) (Fig.~\ref{fig:emojis}C).
By slowly reducing the neighborhood size (e.g., to 20 and 10), the animal emoji cluster in the language model slowly disappears, indicating the face emojis are more similar to the grinning cat than other animal emojis.
The user can change from PCA to t-SNE or UMAP to verify other projection methods also capture this difference.
Interactively hovering over the common objects of the domino and looking at the neighbor's location in the neighborhood plots further reveals that all common neighbors of the grinning cat (a positive sentiment emoji) are found solely in the happy face cluster in the language model.
In contrast, the common neighbors are found in both of image model's clusters, depending on whether the neighbor is a cat or a face.
Together with the user's previous findings, interaction with the dominoes further uncovers how the language embeddings model emoji meaning, and the image embeddings model emoji appearance.

%% file: figures/fig_transfer_learning.tex
\begin{figure*}[t]
    \centering
    \includegraphics[width=\linewidth]{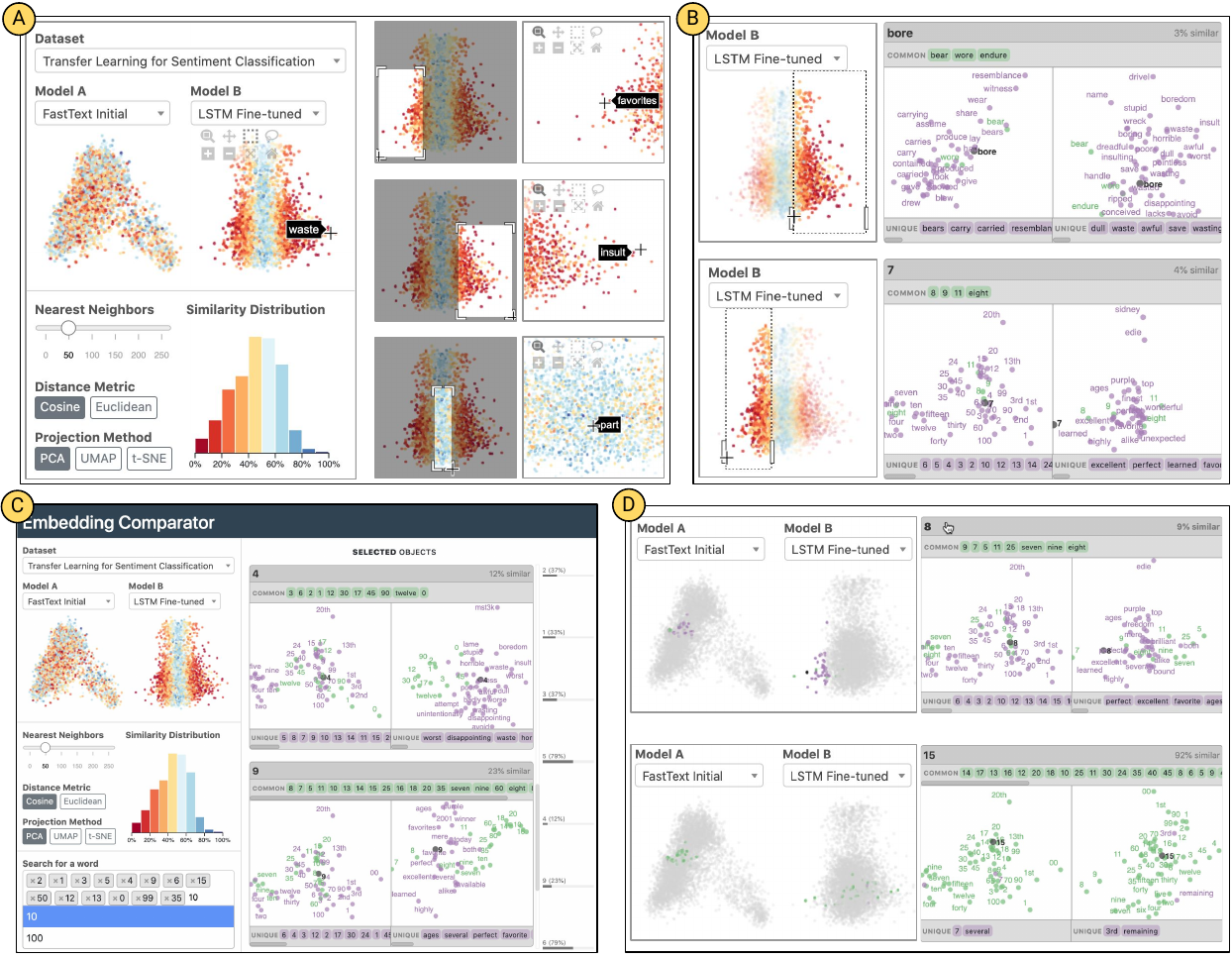}
    \caption{The Embedding Comparator applied to case study \emph{Transfer Learning for Sentiment Classification} compares a word embedding model before and after fine-tuning for sentiment analysis. (A) The global projection plots reveal differences in the structure of the embedding space, such as the fine-tuned model separating negative and positive words. (B) Selecting regions of the global projection plots filters the dominoes, exposing neutral words that have taken on a sentiment meaning as a result of fine-tuning (e.g., \emph{``bore''}). (C) Searching for particular words, such as numbers, shows that numbers in the range 1--10 have shifted in meaning from digits to numerical scoring. (D) Linking back to the global plots shows that this change manifests globally\,---\,while numbers cluster together in the pre-trained embeddings, they are separated in the fine-tuned embeddings.
    }
    \Description[The Embedding Comparator interface being used in the Transfer Learning for Sentiment Classification case study. This figure is visually described in the caption and Section 5.1.]{}
    \label{fig:transfer-learning}
\end{figure*}

%% file: figures/fig_histwords.tex
\begin{figure*}[t]
    \centering
    \includegraphics[width=\linewidth]{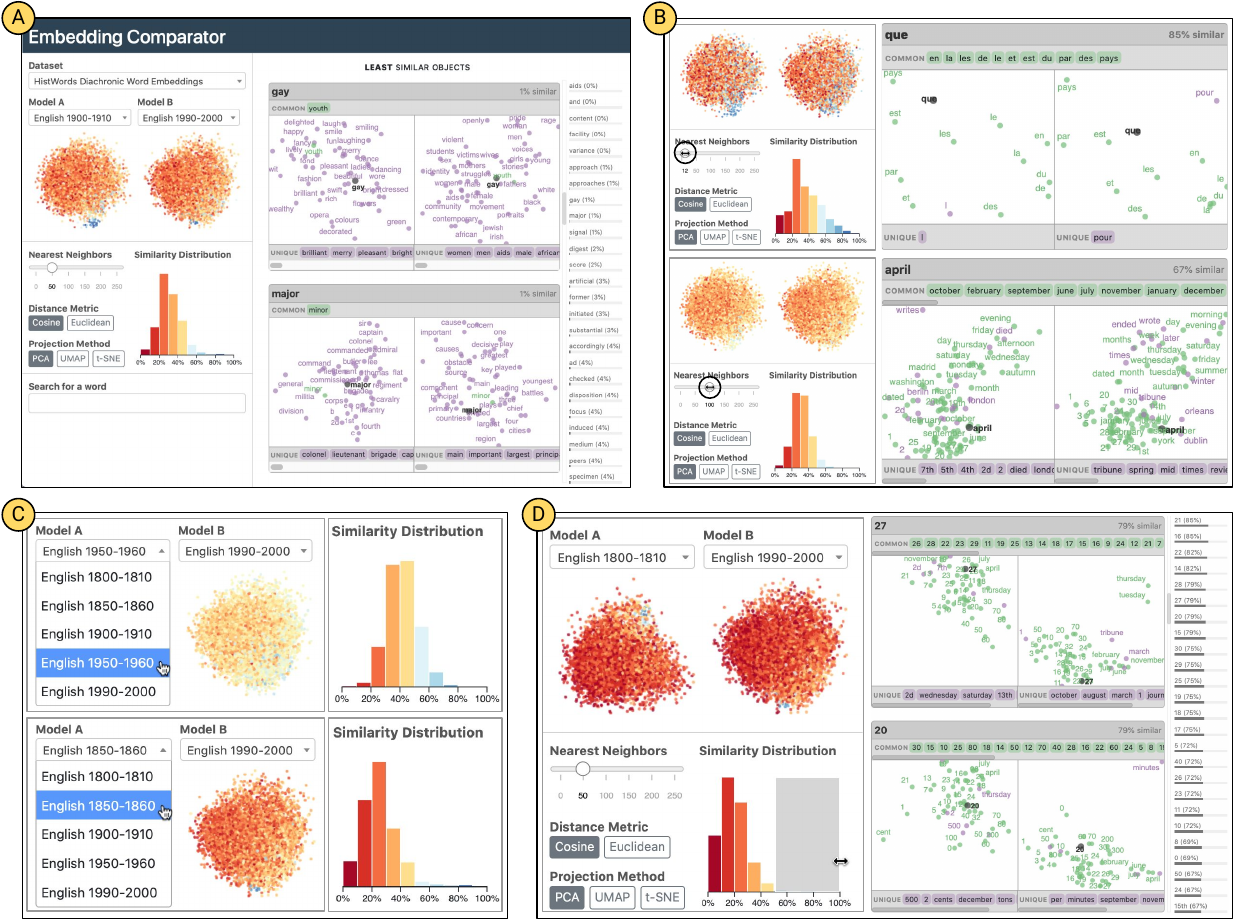}
    \caption{The Embedding Comparator applied to case study \emph{Language Evolution via Diachronic Word Embeddings} compares word embedding models trained on literature from different decades. (A) The sorted domino lists demonstrate how English has changed over time, e.g., the word \emph{``gay''} has changed in meaning from ``happy'' to ``homosexual''.  (B) Varying the $k$-nearest neighbors slider identifies neighborhoods within the embeddings, such as a small cluster of French words and a larger neighborhood of months. (C) Selecting other decades exposes that language similarity decreases as the time between decades increases. (D) Comparing 1800 to 2000 and brushing the right side of the similarity distribution reveals that number usage has remained constant over two centuries.}
    \Description[The Embedding Comparator interface being used in the Language Evolution via Diachronic Word Embeddings case study. This figure is visually described in the caption and Section 5.2.]{}
    \label{fig:hist-words}
\end{figure*}

%% file: figures/fig_emojis.tex
\begin{figure*}[t]
    \centering
    \includegraphics[width=\linewidth]{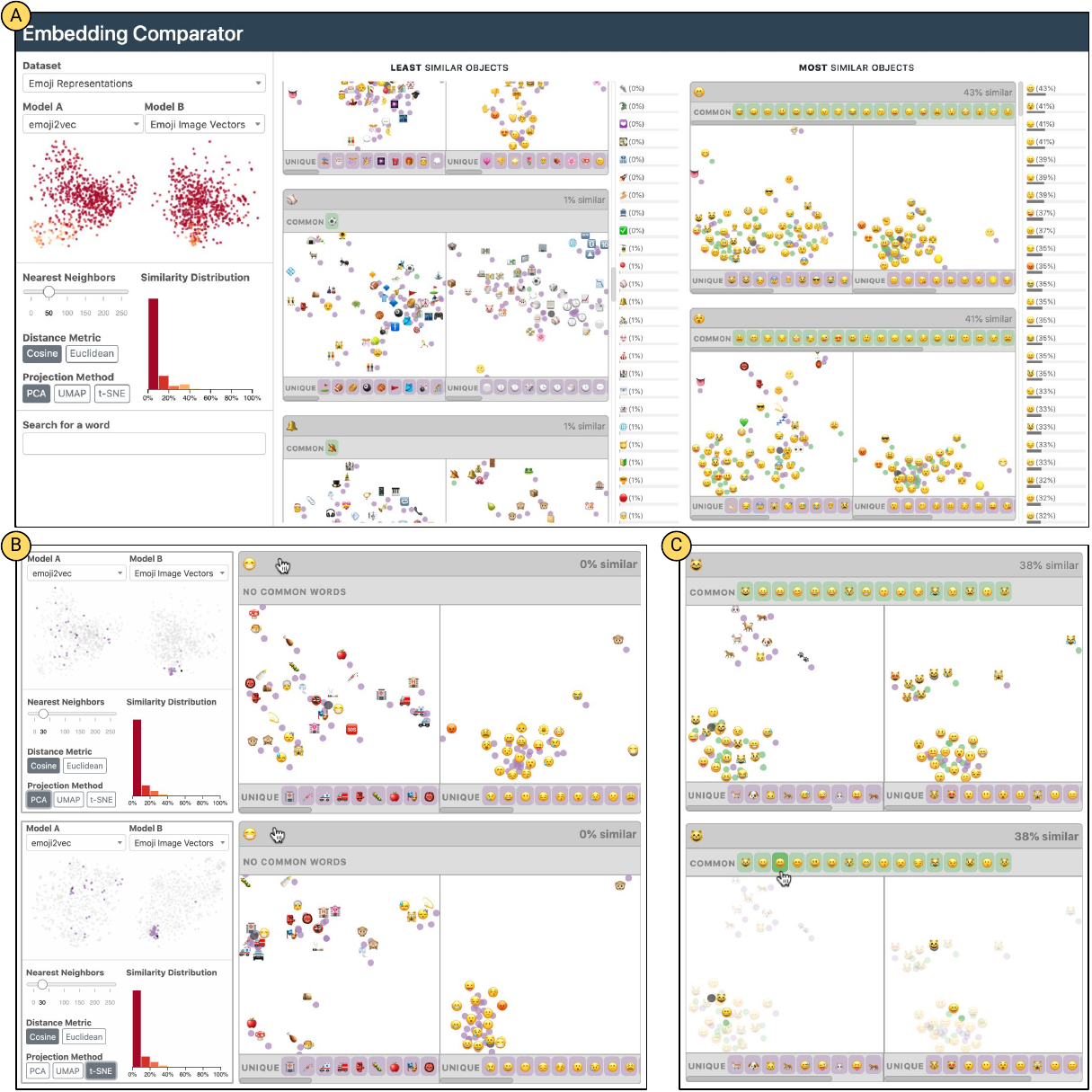}
    \caption{The Embedding Comparator applied to case study:  \emph{Multi-modal Emoji Representations}, compares language embeddings learned from textual descriptions of emojis to image embeddings comprising the emojis' raw pixel values. The language model captures the semantic similarity of emojis, while the image model captures visual similarities. (A) For example, the language embeddings position the baseball emoji with other sports, whereas the image embeddings position it with other white round objects. (B) Hovering on dominoes highlights their position in the global projection plot, revealing a tight cluster of faces in the image model and a more dispersed cluster of medical emojis in the language model. (C) Looking closely at local neighborhoods exposes local clusters, such as face and animal clusters in the language embeddings or cat face and human face clusters in the image embeddings.}
    \Description[The Embedding Comparator interface being used in the Multi-modal Emoji Representations case study. This figure is visually described in the caption and Section 5.3.]{}
    \label{fig:emojis}
\end{figure*}

%% file: sections/06_evaluative_interviews.tex
\section{Evaluation: First-use Studies}
\label{sec:eval-interviews}

The Embedding Comparator was designed to help users efficiently analyze the similarities and differences between embedding spaces.
We performed first-use studies to evaluate its effectiveness.
To simulate users' typical processes (as reported in our formative interviews Section~\ref{sec:formative-interviews}), we had participants compare two embedding spaces using their standard method in a Jupyter Notebook~\cite{kluyver2016jupyter} and with the Embedding Comparator.
In particular, users compared HistWords~\cite{histwords} embeddings from 1800--1810 to embeddings from 1990--2000 (Section~\ref{sec:case-studies-histwords}).
We chose HistWords embeddings because they model English language over time, so no task-specific knowledge was required.
We recruited participants via an open call within our organization.
We performed first-use studies via video chat with 15 representative users: six computer science graduate students, one psychology graduate student, one computational biology graduate student, four ML engineers, and three post-docs.
To ensure representative findings, we recruited users who had experience analyzing and comparing embeddings and self-reported having researched or used embeddings in their work.

Each first-use study lasted 60--90 minutes, and we compensated participants with \$20 Amazon gift cards.
Participants spent half of the study using the Embedding Comparator and half using the Jupyter Notebook.
We randomized the starting interface: seven participants started with the Embedding Comparator, and eight participants started with the Jupyter Notebook.
We began each condition with an introduction to the embeddings and prompted the participants to determine the similarities and differences of the two embeddings spaces.
To enable a fair comparison, in the Jupyter Notebook, we provided code to plot dimensionality reduction projections of the embeddings and print the nearest neighbors of an object.
Further, we encouraged participants to use any tools\,---\,including their code or online GUIs\,---\,that would help them analyze the embeddings.
We asked participants to think aloud throughout the study and, at the end of each condition, to enumerate the similarities and differences between the two embedding spaces and describe what they would do if they had access to the original models and data.

\subsection{Quantitative Results}

\input{figures/fig_quant_results}

We reviewed recordings of each study session and measured the \emph{insights} and \emph{theories} our users identified using each interface.
We define an \emph{insight} as the user indicating that an embedded object, or type of embedded object, is represented similarly or differently in the two embedding spaces in a meaningful way (e.g., ``\textit{gay}'' moving in meaning from ``happy'' to ``homosexual'' or an embedding space projection showing meaningful clusters).
We say the user had a \emph{theory} if they generated a reason for an insight or expressed a desire to look at the original data or models to understand an insight.

We measure the number of insights, number of theories, and time to first insight (measured from the end of our introduction) for both conditions and show participant-level results in Fig.~\ref{fig:quant-results}.
We compute statistical significance using a single-tailed paired sample $t$-test.
Users developed their first insight faster ($p$\,=\,0.003) using the Embedding Comparator ($\mu$\,=\,1:16 min, $\sigma$\,=\,1:20 min) as compared to the Jupyter Notebook ($\mu$\,=\,8:13 min, $\sigma$\,=\,7:54 min).
Users also developed more insights ($p$\,=\,0.004) using the Embedding Comparator ($\mu$\,=\,10.7, $\sigma$\,=\,6.6) as compared to the Jupyter Notebook ($\mu$\,=\,4.1, $\sigma$\,=\,3.3), and users generated more theories ($p$\,=\,0.001) using the Embedding Comparator ($\mu$\,=\,1.7, $\sigma$\,=\,1.6) as compared to the Jupyter Notebook ($\mu$\,=\,0.3, $\sigma$\,=\,0.5).

\subsection{Qualitative Results}

In the Jupyter notebook condition, despite having access to starter code  and being encouraged to use their own code or outside resources, participants struggled to generate meaningful hypotheses that resulted in insights about the embedding spaces.
For example, a common workflow used by nine participants involved generating words they expected to differ between the two spaces and comparing their nearest neighbors.
This process caused a user's biases to drive their exploration (e.g., \emph{``I am looking for a word with two meanings where the meanings are very different''} [P6]; or, \emph{``because I am Arab, I want to look at [the word] Arab''} [P1]).
As a result, users rarely uncovered unexpected results, and they were often frustrated because they had to rely on their intuition\,---\,as P1 said, \emph{``I don't have a sophisticated way of looking through this. Some 1800s and 1900s history knowledge would be useful.''}
Another typical workflow was to evaluate global differences in the embedded representations through summary statistics [P2, P4, P8, P9, P11, P12], cluster analysis [P2, P4, P5, P6], or object ranking [P2, P9, P15].
While these analyses provided insight, they required participants to carefully design experiments that were often tedious to implement.
Using the expressiveness of the Jupyter Notebook, users were able to run experiments that the Embedding Comparator does not support (e.g., computing the average spread of a cluster [P4]).
However, after using the Embedding Comparator, participants expressed that even though it used different mechanisms to compare the embedding spaces, it often provided them the insights they had been trying to surface in the Jupyter Notebook.
For example, P15 said, \emph{``I was trying to do something similar [in the Jupyter Notebook]''}, and P8 expressed, \emph{``This was precisely what I was trying to articulate with the [Jupyter Notebook].''}

When using the Embedding Comparator, users generated insights quickly by looking at the most and least similar lists.
Common insights included the meaning of numbers and religious words (e.g., ``\textit{catholic}'' and ``\textit{god}'') staying constant over time (P1, P2, P5, P7--P13); ``\textit{aids}'' changing in meaning from ``help'' to ``HIV'' (P1, P3, P6, P8, P11--P13); and, ``\textit{logic}'', ``\textit{calculated}'', and ``\textit{volume}'' shifting towards mathematical denotations (P1, P3, P5, P9, P10, P12, P13).
Using the Embedding Comparator, users were able to gain insights into global semantic differences between the embedding spaces. For instance, P9 found \emph{``words that are commonly used without significant changes to their meaning over time are similar between the two embeddings, whereas other words that have changed their usage over time or correspond to a new invention, like the car, are of course dissimilar in the two embeddings. I did not find this in my previous analysis.''}

Users found the dominoes invaluable in their exploration, saying \emph{``it's hard to compress all this information in a Jupyter Notebook, so it's nice to have a tool to be able to browse a lot of words at once''} (P5),  and \emph{``I am able to see [objects] in a graphical view which helps me see [similarities and differences]''} (P3).
Using the dominoes, users were able to analyze many objects at a time and speed up their analysis. 
For instance, P13 found dominoes allowed them to \emph{``get a quick sense of what words distinguish one [embedding space] from the other''} and P1 articulated, \emph{``[the dominoes] allowed me to get a better idea of things because I got more exposure to more words faster''}.

Our first-use studies also validate our local neighborhood similarity metric and the use of this metric to drive our system by sorting dominoes.
Interestingly, without prior knowledge of the Embedding Comparator, one participant implemented a metric similar to ours for computing the local neighborhood intersection during their Jupyter exploration, suggesting local neighborhood similarity is intuitive to users.
However, their implementation was not optimized and did not scale to compute similarity for a sample of more than 40 words, leading to time spent waiting for code to run and uneasiness debugging.
Since the Embedding Comparator calculates similarity for all embedded objects automatically when embeddings are loaded, it shifts users to immediately focus on meaningful parts of the analysis process as opposed to tedious implementation.

Laying out the dominoes ordered by similarity aided users in understanding the similarities and differences between the embedding spaces.
P1 attributed seeing \emph{``a lot more words and what is most similar and least similar''} as the reason why they generated insights faster using the Embedding Comparator and compared the two interfaces by saying \emph{``in the Notebook words were randomly chosen, but here they are sorted, which lets you see things in a more organized way and lets words stick out.''}
P1 found it to be invaluable when generating insights, saying the \emph{``main thing that was useful was having [the Embedding Comparator] order everything, being able to look at the different parts of the distribution, and seeing [objects] in order - the least similar and the most similar.''}
Similarly, P5 reported \emph{``seeing the differences in sets in the nearest neighborhoods is a much easier way to compare the sets of embeddings''} and P2 summarized their experience by saying, \emph{``[the Embedding Comparator] presented the information in a better way and just by doing that we have some sort of a breakthrough.''}

After analyzing the dominoes in the default view of the Embedding Comparator, users varied the parameter controls to further compare models and test hypotheses.
Users often brushed over the similarity distribution to analyze words at a specific similarity value (P1, P5, P7--P15), used the search bar to analyze words that had come to mind during their analysis (P2, P3, P5--P15), and varied $k$ to explore the hierarchy of a word's local neighborhood (P1, P2, P6, P7, P9--P15).
For example, after discovering the neighbors of ``\textit{content}'' were related to ``satisfied'' in one model and ``substances'' in the other, P10 wanted to understand ``how these models represent [words] that have two meanings.''
They hypothesized that both models may have learned both definitions and $k$ was simply too small to see the full neighborhood.
After increasing $k$, P10 found the spaces were still very dissimilar, leading them to theorize one dataset may have contained more scientific text than the other.

When using the Embedding Comparator, users were also able to generate numerous theories including ideas for improving the data representation (e.g., preprocessing numbers into a learned \verb|<num>| token [P5]); new experimental procedures (e.g., training embeddings on native English books vs. books translated from Chinese as a way to compare culture [P7]); reasons for representation failures (e.g., word frequency affects semantic stability over time [P1, P9, P10, P14], or words used in many contexts are susceptible to being represented differently between the two embedding spaces [P2, P10, P12, P13]); and, questions about the original data (\emph{what is the distribution of genres included in the datasets?} [P5--P7, P9, P10, P12] and \emph{what was the data preprocessing pipeline?} [P3, P12]).
After using both interfaces, P5 noted the Embedding Comparator \emph{``made it very clear that they were differences in the embedding spaces''} as compared to the Jupyter Notebook where \emph{``it is a lot harder to find the differences.''}
As P4 said, using the Embedding Comparator \emph{``would be very helpful to my research.''}

%% file: figures/fig_quant_results.tex
\begin{figure}[t]
    \centering
    \hspace{-3mm}
    \begin{subfigure}{.33\linewidth}
      \centering
      \includegraphics[height=1\textwidth]{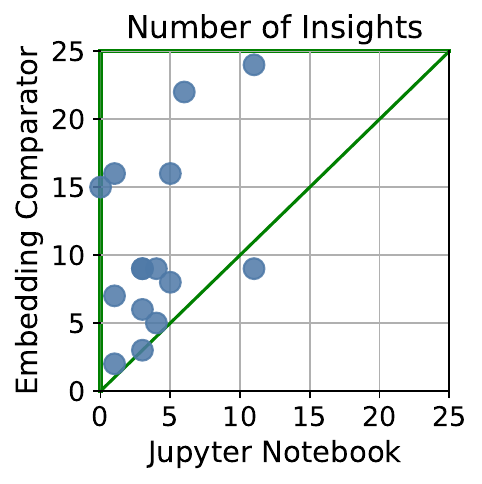}
    \end{subfigure}
    \hspace{-1.7mm}
    \begin{subfigure}{.33\linewidth}
      \centering
      \includegraphics[height=1.0\textwidth]{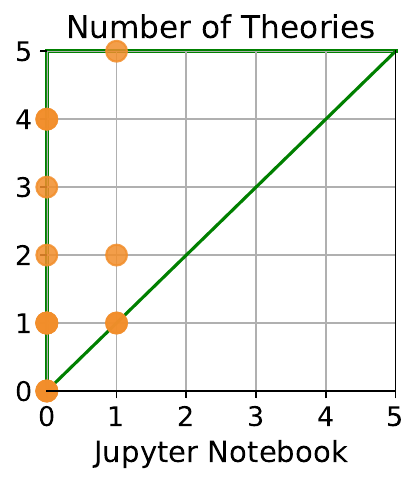}
    \end{subfigure}
    \hspace{-2.75mm}
    \begin{subfigure}{.33\linewidth}
      \centering
      \includegraphics[height=1.0\textwidth]{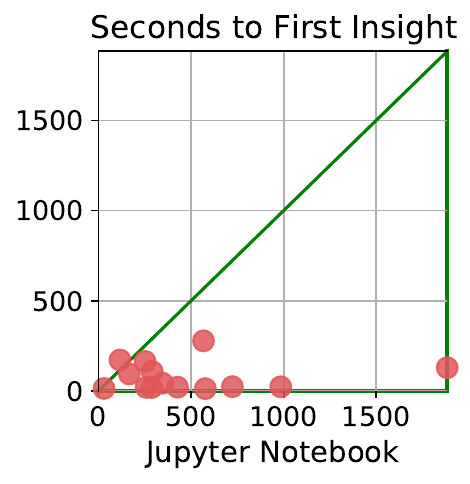}
    \end{subfigure}
    \caption{Using the Embedding Comparator, participants gained more insights and theories and reduced their time to first insight.}
    \Description[First-use study quantitative results. Three scatter plots show the number of insights (left), number of theories (middle), and seconds to first insight (right) during first-use studies by participants using the Embedding Comparator (y-axis) or Jupyter Notebook (x-axis). For number of insights, all points except one lie on or above the diagonal. For number of theories, all points lie on or above the diagonal. For seconds to first insight, all points except one lie on or below the diagonal.]{}
    \label{fig:quant-results}
\end{figure}

%% file: sections/07_conclusion.tex
\section{Discussion and Future Work}
\label{sec:discission}

We present the Embedding Comparator, a novel interactive system for comparing embedding spaces.
Informed by formative interviews conducted with embedding users across disciplines, our design balances between visualizing information about the overall embedding spaces and enabling exploration of local neighborhoods.
To directly surface similarities and differences, a similarity score is computed for every embedded object and encoded across global and local views.
To facilitate rapid comparisons, we introduce \emph{local neighborhood dominoes}: small multiple visualizations of local geometries and lists of common and unique objects.
Interactively linking visualizations of the global embedding space and dominoes permits a tight iterative loop enabling users to alternate between exploring the global space and diving into specific objects of interest.
Through case studies of tasks described by our formative interviewees, we demonstrate how the Embedding Comparator transforms the analysis process from requiring tedious and error-prone manual specification to browsing and interacting with graphical displays.
Moreover, by computing a similarity score for each object, and using it to drive the various views, the Embedding Comparator immediately surfaces interesting insights, and published domain-specific results can be replicated with only a handful of interactions.
High-dimensional datasets also often occur outside of ML, such as from biological gene and protein expression assays used to study cancer~\cite{clarke2008properties} and sensor data used to assess air quality~\cite{engel2012}.  While the case studies presented in our paper are focused on real-world use cases of embeddings in ML described by our interviewees, the Embedding Comparator can be used to compare high-dimensional datasets from other sources.

An important component of our system design was computing a similarity metric for each object across both embedding spaces.
In our interviews, we found users were concerned they may miss unexpected insights because their existing comparison workflows often rely upon ad hoc object selection strategies.
Thus, we chose this metric to systematically prioritize objects that are the most and least similar between the embedding spaces (Design Goal 1), and we encode the similarity scores throughout the global and local views.
While existing systems often require task-specific metrics or alignment of embedding spaces (Section~\ref{sec:related-work}), our metric is agnostic to domain and model and is applicable even when the two embedding spaces stem from different modalities (e.g., natural language and images in Section~\ref{sec:case-studies-emojis}).
In practice, our system surfaced known linguistic insights without requiring task-specific metrics or embedding alignment, while being more extensible to additional tasks described by our interviewees.
We expect that generalizable metrics computed on similarity of local neighborhoods can likewise scaffold future embedding interpretability systems as we demonstrated with the Embedding Comparator.

Another critical aspect of our design was displaying embedded objects as small multiples of local neighborhood dominoes.
While small multiples are a well-understood visualization technique, they remain relatively under-utilized in interpretability systems which largely focus on deeply exploring one input instance at a time.
In contrast, our design is motivated by insights from visualization recommender systems~\cite{wongsuphasawat2015voyager, wongsuphasawat2017voyager}, which have promoted breadth-first exploration of data by adapting Edward Tufte's maxim of prioritizing data variation over design variation~\cite{tufte1983visual}. 
However, a naive application of the small multiples technique can yield an overwhelming experience which burdens users with knowing which small multiple to attend to.
Thus, to bootstrap the exploration process, the Embedding Comparator populates its initial view with two lists of small multiples covering the least similar and most similar objects\,---\,the objects our formative interviewees often described starting their process by examining. 
Future work might consider further iterating on our design by incorporating small multiple summarization techniques such as interactive piling~\cite{lekschas2020generic}. 

The design of the Embedding Comparator suggests several other avenues for compelling future work.
For example, we applied the Embedding Comparator to compare latent embeddings of chemical molecules~\cite{sterling2015zinc} (Appendix~\ref{sec:case-study-chemical-molecules}).
Chemical molecules can be represented as textual SMILES strings or as 2-D structural diagrams.
While the 2-D rendering produces readable dominoes, the length of the SMILES strings limits rapid analysis of the dominoes.
We designed dominoes for concise visual representations of embedded objects; thus, one direction for future work is extracting concise representations from long objects and extending comparison visualizations for long representations.
For instance, for sentences or document embeddings~\cite{le2014distributed}, it would be interesting to explore techniques for identifying specific components of each object (e.g., a subset of words in a document) responsible for the differences in the embeddings and visually display these via dominoes.
Further, in cases where many objects are highly dissimilar (as seen in the chemical molecules example), our LNS metric cannot prioritize specific examples that differ most between the two models.
Future work can explore alternative sorting criteria or enable users to supply domain-specific metrics in these situations.
Finally, our design goals suggest future work visualizing $n$-way model comparisons to accelerate workflows comparing many embedding models and incorporating training data to contextualize an object's similarities or differences across the models.

%% file: sections/08_supplement.tex
\section{Additional Case Studies}
\label{sec:supp-additional-case-studies}

\subsection{Word Embeddings Pre-trained on Different Corpora}
\label{sec:case-studies-twitter}

\input{figures/fig_wiki_twitter}

This case study evokes use cases experienced by both model-driven and data-driven users in our formative interviews: choosing between models that appear equally viable for use in a downstream application (e.g., predicting topics based on customer review text). 
GloVe~\cite{pennington2014glove} is a popular model that offers several variants of embeddings trained with different datasets.
Here, we demonstrate how the Embedding Comparator can be used to understand the impact of training GloVe with data from either Wikipedia \& Newswire or Twitter and replicates known lexical differences between the corpora.

Given the user is tasked with selecting a pretrained model, they begin by analyzing the global shape of each embedding space (Fig.~\ref{fig:wiki-twitter}A).
While the PCA global projection plots show objects continuously distributed in both models, switching to UMAP projections reveals structure and clusters in global embedding spaces.
By hovering over and zooming into the UMAP global projection plots, the user finds clusters of months, names, body parts, and food words within each model.
This finding gives the user confidence that both models have learned meaningful representations of natural language and may be suitable for their task.

Since each model captures meaningful natural language structure, the user is now interested in identifying critical differences between the two models that may help them choose the best model for their task.
The Embedding Comparator immediately reveals a number of differences between these two pre-trained embedding models that arise due to differences in the underlying training data (Fig.~\ref{fig:wiki-twitter}B).
Among the words that differ most are shorthand or slang expressions such as \emph{``bc''}, \emph{``bout''}, and \emph{``def''}.
Scrolling through the local neighborhood dominoes for these words and comparing their unique neighbors reveals specific ways in which their semantic meanings differ between the two models.
For example, in the Wikipedia \& Newswire model, \emph{``def''} is used to mean defeat and \emph{``beats''} (as in sporting results) and hence countries (e.g., \emph{``canada''} and \emph{``usa''}), whereas in the Twitter model, \emph{``def''} relates to conversational words such as \emph{``definitely''} and \emph{``probably''}.
Another insight revealed by our system is the difference in languages present in the training corpora.
Words such as \textit{``era''}, \textit{``dale''}, and \textit{``solo''} take on their English meanings in the Wikipedia \& Newswire model, but are related to Spanish words in the Twitter model.
This finding suggests that the Twitter model was trained on multi-lingual text, while the Wikipedia \& Newswire model may have been trained solely on English text.
Finally, the Embedding Comparator reveals how words such as \emph{``swift''} and \emph{``galaxy''} may be used very differently in different media.
In Wikipedia \& Newswire, \emph{``swift''} refers to the adjective swift (i.e., quick), whereas on Twitter, \emph{``swift''} refers to the musical artist Taylor Swift.
Likewise, \emph{``galaxy''} refers either to space or to Samsung Galaxy electronics based on whether embeddings were trained on Wikipedia \& Newswire or on Twitter, respectively.

Lexical comparison between Wikipedia and Twitter via embeddings has been explored in previous work~\cite{tan2015lexical}.
For data-driven users, the Embedding Comparator surfaces characteristic words previously reported to differ most between the two corpora, including \emph{``bc''} and \emph{``ill''} (Fig.~\ref{fig:wiki-twitter}B).
Moreover, these words are immediately surfaced as a result of our similarity metric without requiring alignment of the two embedding spaces or introducing variance by learning a linear transformation using a stochastic procedure.
Using these insights from the Embedding Comparator, a user can make a more informed decision about which set of embeddings may be more appropriate to adopt for their system.
For example, if classifying longer or more formal customer reviews, the model trained on Wikipedia \& Newswire would likely perform better, but if classifying casually written reviews that contain slang or multi-lingual text, the Twitter corpus may generalize better to real-world data.

\subsection{Chemical Molecules}
\label{sec:case-study-chemical-molecules}

\input{figures/fig_chemical_molecules}

We applied the Embedding Comparator to compare latent embeddings of chemical molecules~\cite{sterling2015zinc} learned by two different variational autoencoder (VAE) architectures, Grammar VAE~\cite{kusner2017grammar} and junction tree VAE~\cite{jin2018junction} (Fig.~\ref{fig:chemical-molecules}).
This application demonstrates a potential limitation of our system on long object labels.
In Fig.~\ref{fig:chemical-molecules}A, the molecules are represented as textual SMILES strings~\cite{weininger1988smiles}, which are often used to describe chemical structures.
We find that these string representations are long, obfuscating the ability to rapidly scroll through dominoes.
Fig.~\ref{fig:chemical-molecules}B shows the same models but where a 2-D structure of each molecule is shown instead, rendering dominoes more readable.

\section{Case Study Details}
\label{sec:supp-case-studies-details}

Here, we detail the datasets and preprocessing steps used in our case studies.

\subsection{Transfer Learning for Sentiment Classification}
We downloaded pre-trained fastText~\cite{mikolov2018advances} word embeddings, which are available online at \url{https://fasttext.cc/docs/en/english-vectors.html}.
We use the 300-dimensional \verb|wiki-news-300d-1M| embeddings consisting of 1 million word vectors trained on Wikipedia 2017, UMBC webbase corpus, and statmt.org news datasets.

We train an LSTM~\cite{hochreiter1997long} to classify binary sentiment in movie reviews from the Large Movie Review dataset~\cite{imdbdataset} containing 25000 training reviews and 25000 test reviews from the Internet Movie Database (IMDb).
We use default tokenization settings for this dataset as provided in Keras~\cite{chollet2015keras}.
We define our vocabulary as the top 5000 most frequent words in the movie review dataset and truncate reviews to a maximum length of 500 words (with pre-padding).
Our recurrent neural network architecture is defined as follows:
\begin{enumerate}
    \item \textbf{Input/Embeddings Layer}: Sequence with 500 words. The word at each timestep is represented by a 300-dimensional embedding.
    \item \textbf{LSTM}: Recurrent layer with 100-unit LSTM (forward direction only, dropout = 0.2, recurrent dropout = 0.2).
    \item \textbf{Dense}: 1 neuron (sentiment output), sigmoid activation.
\end{enumerate}
Prior to training, the embeddings are initialized using the pre-trained fastText embeddings.
Of our vocabulary of size 5000, 4891 tokens were present in the fastText embeddings.
For tokens not present in fastText, we initialize embeddings as all-zero vectors.

We train our model for 3 epochs (batch size = 64) with the Adam optimizer~\cite{adam} using default parameters in Keras~\cite{chollet2015keras} to minimize binary cross-entropy on the training set.
The final model achieves 84.7\% test set accuracy (85.4\% training set accuracy).
We did not further tune the architecture or hyperparameters.

For analysis in the Embedding Comparator, we output the initial fastText embeddings and fine-tuned embeddings for the 4891 words whose embeddings were initialized from fastText.

\subsection{Language Evolution via Diachronic Word Embeddings}
We use the HistWords dataset~\cite{histwords} in our case study of diachronic word embeddings, which have been shown to exhibit changes in semantic meaning of words over time.
We use pre-trained word embeddings from~\cite{histwords}, accessed at \url{https://nlp.stanford.edu/projects/histwords/}.
We use the \verb|All English (1800s-1990s)| set of embeddings, which are 300-dimensional word2vec embeddings~\cite{mikolov2013distributed}.
This dataset provides word embeddings trained on English books from each decade from 1800 to 2000.
For exploration in the Embedding Comparator, we select embeddings taken from five different decades spanning this time period: 1800-1810, 1850-1860, 1900-1910, 1950-1960, and 1990-2000.
We filter each embedding space to the top 10000 most frequent words from its decade and compute the intersection of these sets over the five decades we selected, producing a vocabulary containing 6121 words from each model for comparison in the Embedding Comparator.

\subsection{Multimodal Emoji Representations}
We use pre-trained 300-dimensional emoji embeddings from \\emoji2vec~\cite{eisner2016emoji2vec}, accessed from \url{https://github.com/uclnlp/emoji2vec}.
Our image (pixel) embedding model uses emoji images obtained from \url{https://github.com/iamcal/emoji-data/blob/master/sheet_apple_16.png}.
The raw $18 \times 18 \times 4$ RGBA images are flattened into a 1296-dimensional vector, which is the pixel model embedding.
We use 670 total emojis that are common across these datasets.

\subsection{Word Embeddings Pre-trained on Different Corpora}
In this case study, we use pre-trained embeddings from GloVe~\cite{pennington2014glove}, available online at \url{https://nlp.stanford.edu/projects/glove/}.
The Wikipedia/newswire embeddings were trained on the Wikipedia 2014 and Gigaword 5 (newswire text) datasets containing 6 billion tokens (GloVe 6B), while the Twitter word embeddings were trained on text from Twitter tweets containing 27 billion tokens (GloVe 27B).
We use the 100-dimensional embeddings trained on each of these corpora.
We filter each of the embedding models to the top 10K most frequent words from its respective corpus and then intersect the resulting vocabularies, giving a shared vocabulary containing 3303 words.
We use the Embedding Comparator to compare embeddings from each model for words in this shared vocabulary.

\section{Additional Dominoes}
\label{sec:supp-gallery}

\input{figures/gallery}

%% file: figures/fig_wiki_twitter.tex
\begin{figure*}[ht]
    \centering
    \includegraphics[width=\linewidth]{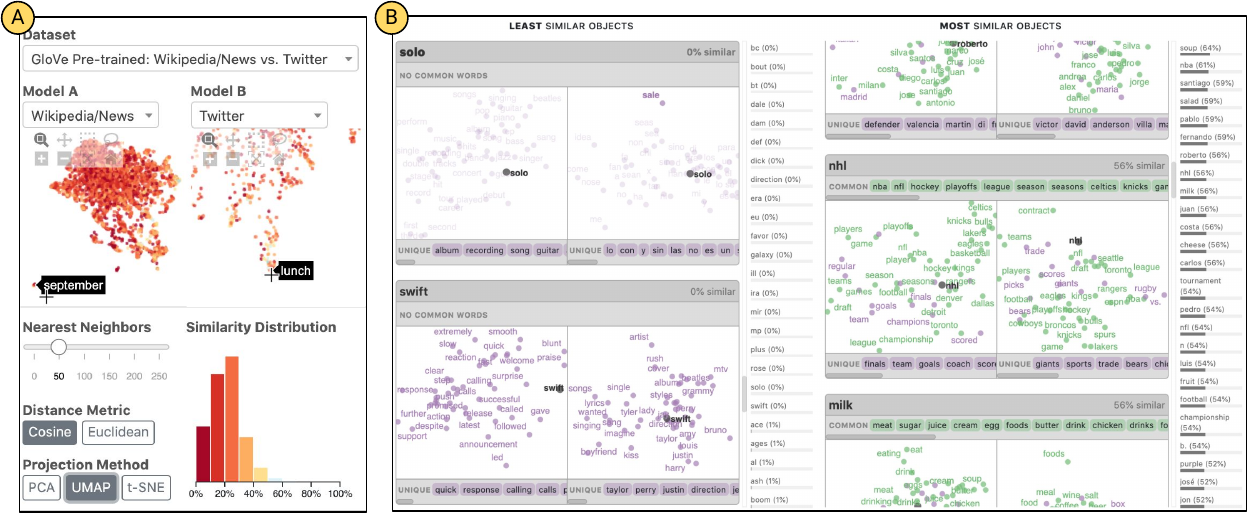}
    \caption{The Embedding Comparator, applied to case study: \emph{Word Embeddings Pre-trained on Different Corpora}, compares a word embedding model trained on Wikipedia and Newswire text to a model trained on Twitter text. (A) Despite using the same architecture, the similarity distribution and global projection plots show the models represent words differently. (B) Looking at the dominoes suggests the Wikipedia/Newswire model was trained on proper English text, while the Twitter model contains Spanish words (e.g., \emph{``solo''}) and emphasizes popular culture references (e.g., \emph{``swift''}).}
    \Description[The Embedding Comparator interface being used in the Word Embeddings Pre-trained on Different Corpora case study. This figure is visually described in the caption and Section A.1.]{}
    \label{fig:wiki-twitter}
\end{figure*}

%% file: figures/fig_chemical_molecules.tex
\begin{figure*}[ht]
    \centering
    \includegraphics[width=\linewidth]{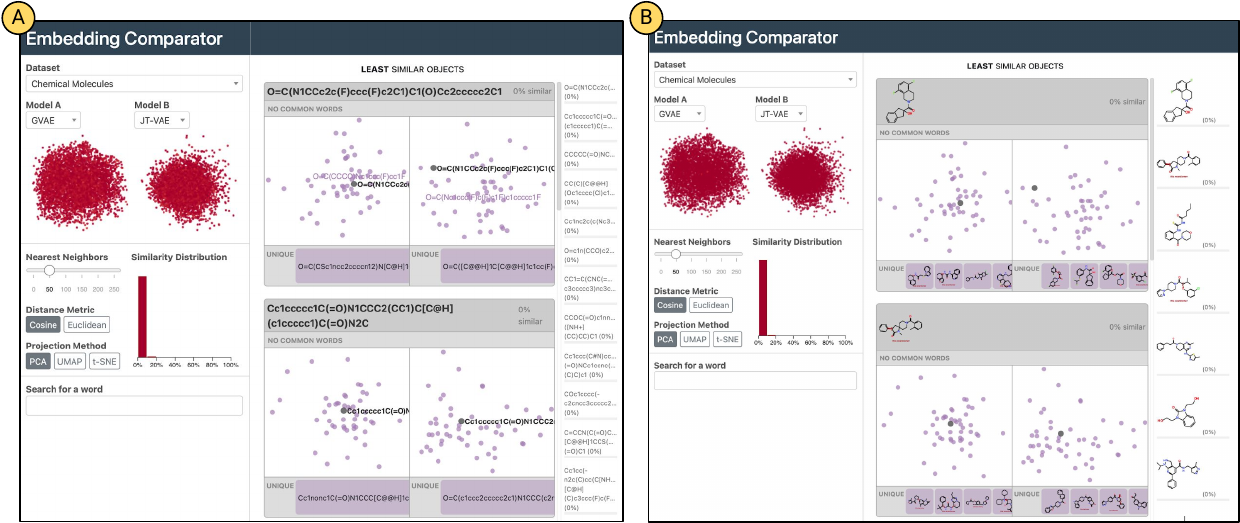}
    \caption{View of the Embedding Comparator applied to two embeddings of chemical molecules learned by different variational autoencoder architectures. (A) Molecules are represented as SMILES strings. (B) Molecules are represented as 2-D structures.}
    \Description[The Embedding Comparator interface being used to compare chemical molecule embeddings. This figure is visually described in the caption and Section A.2.]{}
    \label{fig:chemical-molecules}
\end{figure*}

%% file: figures/gallery.tex
\input{figures/fig_gallery_transfer_learning}
\input{figures/fig_gallery_histwords}
\input{figures/fig_gallery_emoji}
\input{figures/fig_gallery_twitter_wiki}
\input{figures/fig_gallery_similar}

%% file: figures/fig_gallery_transfer_learning.tex
\begin{figure*}[ht]
    \centering
    \includegraphics[width=\linewidth]{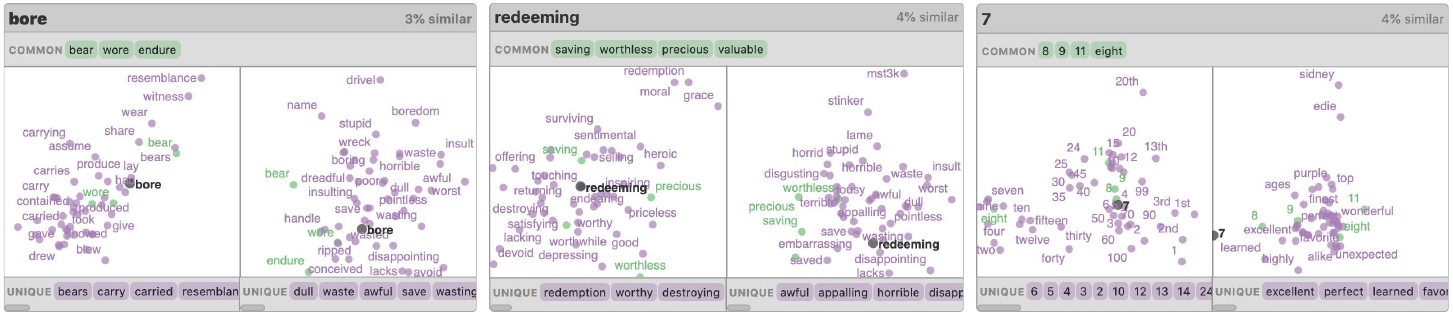}
    \caption{Additional dominoes from case study: \emph{Transfer Learning for Sentiment Classification}. The word \emph{``bore''} has changed in meaning from a general definition: ``carried'', to a more sentiment rich definition: ``dull''. \emph{``redeeming''} has changed from the positive sentiment definition: ``compensate for faults'' to a negative sentiment definition likely related to the reviewer idiom ``no redeeming qualities''. The number \emph{``7''} has changed from its definition as a numeric symbol to a number indicative of score (e.g., 7 out of 10).}
    \label{fig:gallery-transfer-learning}
    \Description[Three dominoes from the Transfer Learning for Sentiment Classification case study. The leftmost domino represents the word "bore". Its common neighbors are "bear", "wore", and "endure". The three closest unique neighbors in the fastText embeddings are "bears", "carry", and "carried". The three closest unique neighbors in the fine-tuned embeddings are "dull", "waste", and "awful". The center domino represents the word "redeeming". Its common neighbors are "saving", "worthless", "precious", and "valuable". The three closest unique neighbors in the fastText embeddings are "redemption", "worthy", and "destroying". The three closest unique neighbors in the fine-tuned embeddings are "awful", "appalling", and "horrible". The rightmost domino represents the word "7". Its common neighbors are "8", "9", "11", and "eight". The three closest unique neighbors in the fastText embeddings are "6", "5", and "4". The three closest unique neighbors in the fine-tuned embeddings are "excellent", "perfect", and "learned".]{}
\end{figure*}

%% file: figures/fig_gallery_histwords.tex
\begin{figure*}[ht]
    \centering
    \includegraphics[width=\linewidth]{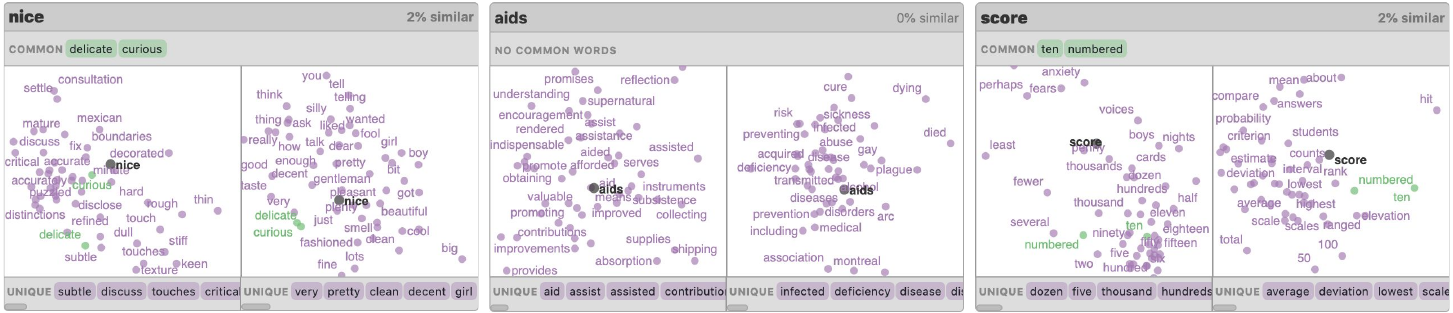}
    \caption{Additional dominoes from case study: \emph{Language Evolution via Diachronic Word Embeddings}. The domino for \emph{``nice''} shows its definition change from ``fine'' in 1800-1810 to ``pleasant'' in 1900--1910. The word \emph{``aids''} was synonymous with ``assists'' in 1900--1910 but later became associated with HIV/AIDS in 1990--2000.  Over the course of the 20th century, \emph{``score''} changed from a measure of time (e.g., four score) to a measure of rank.}
    \label{fig:gallery-hist-words}
    \Description[Three dominoes from the Language Evolution via Diachronic Word Embeddings case study. The leftmost domino represents the word "nice". Its common neighbors are "delicate" and "curious". The three closest unique neighbors in the 1800s embeddings are "subtle", "discuss", and "touches". The three closest unique neighbors in the 1900s embeddings are "very", "pretty", and "clean". The center domino represents the word "aids". Its has no common neighbors. The three closest unique neighbors in the 1800s embeddings are "aid", "assist", and "assisted". The three closest unique neighbors in the 1900s embeddings are "infected", "deficiency", and "disease". The rightmost domino represents the word "score". Its common neighbors are "ten" and "numbered". The three closest unique neighbors in the 1800s embeddings are "dozen", "five", and "average". The three closest unique neighbors in the 1900s embeddings are "average", "deviation", and "lowest".]{}
\end{figure*}

%% file: figures/fig_gallery_emoji.tex
\begin{figure*}[ht]
    \centering
    \includegraphics[width=\linewidth]{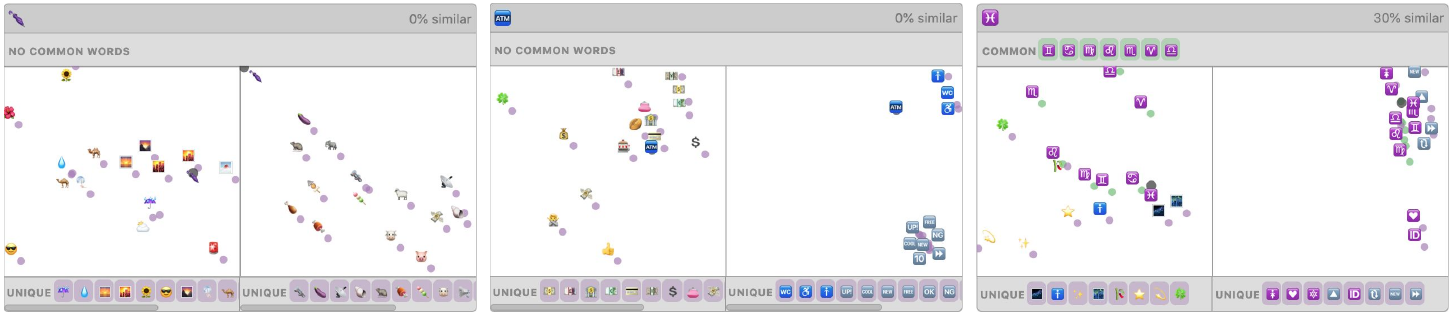}
    \caption{Additional dominoes from case study: \emph{Multimodal Emoji Representations}. In the language model, the umbrella emoji is related to other weather emojis, whereas in the image model it is related to other slanted emojis. The ATM emoji is related to money emojis in the language model and related to other blue square emojis in the image model. The Pisces emoji is related to other astrology signs in both models because they share astrological meaning and are visually similar.}
    \label{fig:gallery-emoji}
    \Description[Three dominoes from the Multimodal Emoji Representations case study. The leftmost domino represents umbrella emoji. Its has no common neighbors. The three closest unique neighbors in the language embeddings are the umbrella with rain drops, droplet, and sunrise emojis. The three closest unique neighbors in the image embeddings are the nut and bolt, eggplant, and satellite antenna emojis. The center domino represents the ATM emoji. Its has no common neighbors. The three closest unique neighbors in the language embeddings are the dollar banknote, pound banknote, and bank emojis. The three closest unique neighbors in the image embeddings are the water closet, wheelchair symbol, and men's room emojis. The rightmost domino represents the Pisces emoji. Its three closest common neighbors are Gemini, Cancer, and Virgo emojis. The three closest unique neighbors in the language embeddings are the milky way, men's room, and sparkles emojis. The three closest unique neighbors in the image embeddings are the woman's room, heart decoration, and dotted six-pointed star emojis.]{}
\end{figure*}

%% file: figures/fig_gallery_twitter_wiki.tex
\begin{figure*}[ht]
    \centering
    \includegraphics[width=\linewidth]{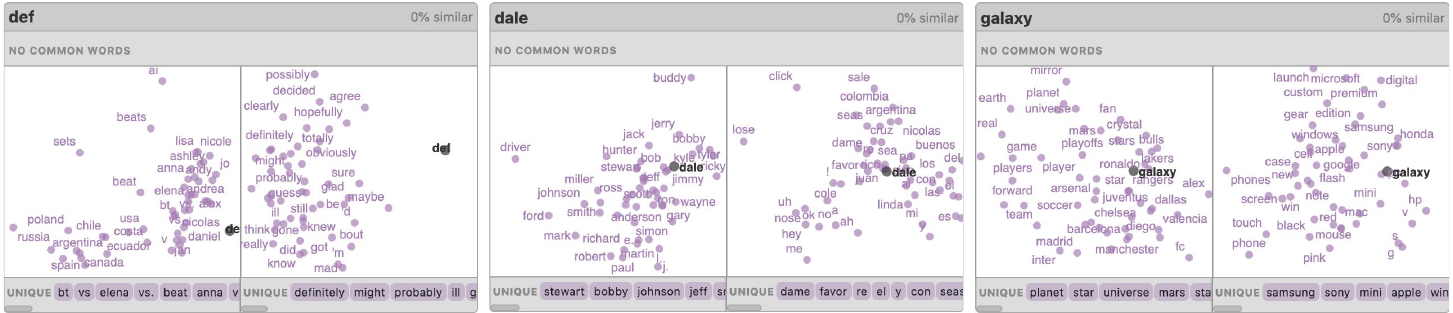}
    \caption{Additional dominoes from case study: \emph{Word Embeddings Pre-trained on Different Corpora}. Using the model trained on news text \emph{``def''} is short for ``defeated'', whereas using the model trained on Twitter data \emph{``def''} is slang for ``definitely''. The word \emph{``dale''} is an English name in the news model, but is a represented by its Spanish meaning in the Twitter model. \emph{``galaxy''} in the news model is related to space, but in the Twitter model is related to the Samsung Galaxy line of phones.}
    \label{fig:gallery-twitter-wiki}
    \Description[Three dominoes from the Word Embeddings Pre-trained on Different Corpora case study. The leftmost domino represents the word "def". Its has no common neighbors. The three closest unique neighbors in the news embeddings are "bt", "vs", and "elena". The three closest unique neighbors in the Twitter embeddings are "definitely", "might", and "probably". The center domino represents the word "dale". Its has no common neighbors. The three closest unique neighbors in the news embeddings are "stewart", "bobby", and "johnson". The three closest unique neighbors in the Twitter embeddings are "dame", "favor", and "re". The rightmost domino represents the word "galaxy". Its has no common neighbors. The three closest unique neighbors in the news embeddings are "planet", "star", and "universe". The three closest unique neighbors in the Twitter embeddings are "samsung", "sony", and "mini".]{}
\end{figure*}

%% file: figures/fig_gallery_similar.tex
\begin{figure*}[ht]
    \centering
    \includegraphics[width=\linewidth]{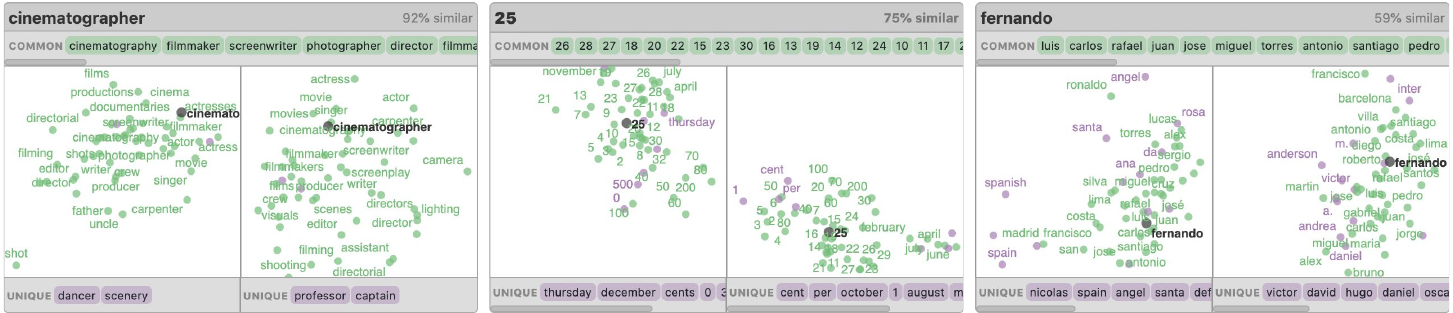}
    \caption{Additional dominoes where the neighborhood of the word has not changed. From the \emph{Transfer Learning for Sentiment Classification} case study, \emph{``cinematographer''} does not change when fine tuning a general English model on a movie review dataset because \emph{``cinematographer''} already refers to the film industry. From the \emph{Language Evolution via Diachronic Word Embeddings} case study, \emph{``25''} has not changed from 1800 to 2000, indicating that numbers are not susceptible to chronological changes in meaning. \emph{``fernando''} from the \emph{Word Embeddings Pre-trained on Different Corpora} case study does not change in meaning whether the model was trained on news data or Twitter data likely because it is a proper noun with no other meanings.}
    \Description[Three similar dominoes from the Embedding Comparator case studies. The leftmost domino represents the word "cinematographer" from the Transfer Learning for Sentiment Classification case study. Its three closest common neighbors are "cinematography", "filmmaker", and "screenwriter". The unique neighbors in the fastText embeddings are "dancer" and "scenery". The unique neighbors in the fine-tuned embeddings are "professor" and "captain". The center domino represents the word "25" from the Language Evolution via Diachronic Word Embeddings case study. Its three closest common neighbors are "26", "28", and "27". The three closest unique neighbors in the 1800s embeddings are "thursday", "december", and "cents". The three closest unique neighbors in the 2000s embeddings are "cent", "per", and "october". The rightmost domino represents the word "fernando" from the Word Embeddings Pre-trained on Different Corpora case study. Its three closest common neighbors are "luis", "carlos" and "rafael". The three closest unique neighbors in the news embeddings are "nicolas", "spain", and "angel". The three closest unique neighbors in the Twitter embeddings are "victor", "david", and "hugo".]{}
    \label{fig:gallery-similar}
\end{figure*}

%% file: embedding_comparator.bbl

\begin{thebibliography}{78}


\ifx \showCODEN    \undefined \def \showCODEN     #1{\unskip}     \fi
\ifx \showDOI      \undefined \def \showDOI       #1{#1}\fi
\ifx \showISBNx    \undefined \def \showISBNx     #1{\unskip}     \fi
\ifx \showISBNxiii \undefined \def \showISBNxiii  #1{\unskip}     \fi
\ifx \showISSN     \undefined \def \showISSN      #1{\unskip}     \fi
\ifx \showLCCN     \undefined \def \showLCCN      #1{\unskip}     \fi
\ifx \shownote     \undefined \def \shownote      #1{#1}          \fi
\ifx \showarticletitle \undefined \def \showarticletitle #1{#1}   \fi
\ifx \showURL      \undefined \def \showURL       {\relax}        \fi
\providecommand\bibfield[2]{#2}
\providecommand\bibinfo[2]{#2}
\providecommand\natexlab[1]{#1}
\providecommand\showeprint[2][]{arXiv:#2}

\bibitem[\protect\citeauthoryear{Allen and Hospedales}{Allen and
  Hospedales}{2019}]%
        {allen2019analogies}
\bibfield{author}{\bibinfo{person}{Carl Allen} {and}
  \bibinfo{person}{Timothy~M. Hospedales}.} \bibinfo{year}{2019}\natexlab{}.
\newblock \showarticletitle{Analogies Explained: Towards Understanding Word
  Embeddings}. In \bibinfo{booktitle}{\emph{Proceedings of the International
  Conference on Machine Learning ({ICML})}}, Vol.~\bibinfo{volume}{97}.
  \bibinfo{publisher}{{PMLR}}, \bibinfo{address}{Long Beach, {USA}},
  \bibinfo{pages}{223--231}.
\newblock


\bibitem[\protect\citeauthoryear{Arendt, Nur, Huang, Fair, and Dou}{Arendt
  et~al\mbox{.}}{2020}]%
        {arendt2020parallel}
\bibfield{author}{\bibinfo{person}{Dustin~L. Arendt}, \bibinfo{person}{Nasheen
  Nur}, \bibinfo{person}{Zhuanyi Huang}, \bibinfo{person}{Gabriel Fair}, {and}
  \bibinfo{person}{Wenwen Dou}.} \bibinfo{year}{2020}\natexlab{}.
\newblock \showarticletitle{Parallel Embeddings: a Visualization Technique for
  Contrasting Learned Representations}. In
  \bibinfo{booktitle}{\emph{Proceedings of the International Conference on
  Intelligent User Interfaces ({IUI})}}. \bibinfo{publisher}{{ACM}},
  \bibinfo{address}{Cagliari, Italy}, \bibinfo{pages}{259--274}.
\newblock


\bibitem[\protect\citeauthoryear{Bau, Zhou, Khosla, Oliva, and Torralba}{Bau
  et~al\mbox{.}}{2017}]%
        {netdissect2017}
\bibfield{author}{\bibinfo{person}{David Bau}, \bibinfo{person}{Bolei Zhou},
  \bibinfo{person}{Aditya Khosla}, \bibinfo{person}{Aude Oliva}, {and}
  \bibinfo{person}{Antonio Torralba}.} \bibinfo{year}{2017}\natexlab{}.
\newblock \showarticletitle{Network Dissection: Quantifying Interpretability of
  Deep Visual Representations}. In \bibinfo{booktitle}{\emph{Proceedings of the
  Conference on Computer Vision and Pattern Recognition ({CVPR})}}.
  \bibinfo{publisher}{{IEEE}}, \bibinfo{address}{Honolulu, {USA}},
  \bibinfo{pages}{3319--3327}.
\newblock


\bibitem[\protect\citeauthoryear{Bau, Zhu, Strobelt, Zhou, Tenenbaum, Freeman,
  and Torralba}{Bau et~al\mbox{.}}{2019}]%
        {bau2019gandissect}
\bibfield{author}{\bibinfo{person}{David Bau}, \bibinfo{person}{Jun{-}Yan Zhu},
  \bibinfo{person}{Hendrik Strobelt}, \bibinfo{person}{Bolei Zhou},
  \bibinfo{person}{Joshua~B. Tenenbaum}, \bibinfo{person}{William~T. Freeman},
  {and} \bibinfo{person}{Antonio Torralba}.} \bibinfo{year}{2019}\natexlab{}.
\newblock \showarticletitle{{GAN} Dissection: Visualizing and Understanding
  Generative Adversarial Networks}. In \bibinfo{booktitle}{\emph{Proceedings of
  the International Conference on Learning Representations ({ICLR})}}.
  \bibinfo{publisher}{OpenReview.net}, \bibinfo{address}{New Orleans, USA}.
\newblock


\bibitem[\protect\citeauthoryear{Bengio and Heigold}{Bengio and
  Heigold}{2014}]%
        {bengio2014word}
\bibfield{author}{\bibinfo{person}{Samy Bengio} {and} \bibinfo{person}{Georg
  Heigold}.} \bibinfo{year}{2014}\natexlab{}.
\newblock \showarticletitle{Word Embeddings for Speech Recognition}. In
  \bibinfo{booktitle}{\emph{Proceedings of the Conference of the International
  Speech Communication Association ({INTERSPEECH})}}.
  \bibinfo{publisher}{{ISCA}}, \bibinfo{address}{Singapore},
  \bibinfo{pages}{1053--1057}.
\newblock


\bibitem[\protect\citeauthoryear{Bepler and Berger}{Bepler and Berger}{2019}]%
        {bepler2018learning}
\bibfield{author}{\bibinfo{person}{Tristan Bepler} {and}
  \bibinfo{person}{Bonnie Berger}.} \bibinfo{year}{2019}\natexlab{}.
\newblock \showarticletitle{Learning protein sequence embeddings using
  information from structure}. In \bibinfo{booktitle}{\emph{Proceedings of the
  International Conference on Learning Representations ({ICLR})}}.
  \bibinfo{publisher}{OpenReview.net}, \bibinfo{address}{New Orleans, {USA}}.
\newblock


\bibitem[\protect\citeauthoryear{Bileschi, Belanger, Bryant, Sanderson, Carter,
  Sculley, Bateman, DePristo, and Colwell}{Bileschi et~al\mbox{.}}{2022}]%
        {bileschi2019using}
\bibfield{author}{\bibinfo{person}{Maxwell~L Bileschi}, \bibinfo{person}{David
  Belanger}, \bibinfo{person}{Drew~H Bryant}, \bibinfo{person}{Theo Sanderson},
  \bibinfo{person}{Brandon Carter}, \bibinfo{person}{D Sculley},
  \bibinfo{person}{Alex Bateman}, \bibinfo{person}{Mark~A DePristo}, {and}
  \bibinfo{person}{Lucy~J Colwell}.} \bibinfo{year}{2022}\natexlab{}.
\newblock \showarticletitle{Using deep learning to annotate the protein
  universe}.
\newblock \bibinfo{journal}{\emph{Nature Biotechnology}}
  (\bibinfo{year}{2022}).
\newblock


\bibitem[\protect\citeauthoryear{Carroll and Pirolli}{Carroll and
  Pirolli}{2003}]%
        {pirolli2003exploring}
\bibfield{author}{\bibinfo{person}{John Carroll} {and} \bibinfo{person}{Peter
  Pirolli}.} \bibinfo{year}{2003}\natexlab{}.
\newblock \bibinfo{booktitle}{\emph{HCI Models, Theories and Frameworks}}.
\newblock \bibinfo{publisher}{Morgan Kaufmann}, \bibinfo{address}{San
  Francisco, CA}, Chapter Exploring and Finding Information,
  \bibinfo{pages}{157–191}.
\newblock


\bibitem[\protect\citeauthoryear{Carter, Mueller, Jain, and Gifford}{Carter
  et~al\mbox{.}}{2019}]%
        {carter2018made}
\bibfield{author}{\bibinfo{person}{Brandon Carter}, \bibinfo{person}{Jonas
  Mueller}, \bibinfo{person}{Siddhartha Jain}, {and} \bibinfo{person}{David~K.
  Gifford}.} \bibinfo{year}{2019}\natexlab{}.
\newblock \showarticletitle{What made you do this? Understanding black-box
  decisions with sufficient input subsets}. In
  \bibinfo{booktitle}{\emph{Proceedings of the International Conference on
  Artificial Intelligence and Statistics ({AISTATS})}}.
  \bibinfo{publisher}{{PMLR}}, \bibinfo{address}{Naha, Japan},
  \bibinfo{pages}{567--576}.
\newblock


\bibitem[\protect\citeauthoryear{Chen, Tao, and Lin}{Chen
  et~al\mbox{.}}{2018}]%
        {chen2018visual}
\bibfield{author}{\bibinfo{person}{Juntian Chen}, \bibinfo{person}{Yubo Tao},
  {and} \bibinfo{person}{Hai Lin}.} \bibinfo{year}{2018}\natexlab{}.
\newblock \showarticletitle{Visual Exploration and Comparison of Word
  Embeddings}.
\newblock \bibinfo{journal}{\emph{Journal of Visual Languages \& Computing}}
  \bibinfo{volume}{48} (\bibinfo{year}{2018}), \bibinfo{pages}{178--186}.
\newblock


\bibitem[\protect\citeauthoryear{Chollet et~al\mbox{.}}{Chollet
  et~al\mbox{.}}{2015}]%
        {chollet2015keras}
\bibfield{author}{\bibinfo{person}{Fran\c{c}ois Chollet} {et~al\mbox{.}}}
  \bibinfo{year}{2015}\natexlab{}.
\newblock \bibinfo{title}{Keras}.
\newblock \bibinfo{howpublished}{\url{https://keras.io}}.
\newblock


\bibitem[\protect\citeauthoryear{Clarke, Ressom, Wang, Xuan, Liu, Gehan, and
  Wang}{Clarke et~al\mbox{.}}{2008}]%
        {clarke2008properties}
\bibfield{author}{\bibinfo{person}{Robert Clarke}, \bibinfo{person}{Habtom~W
  Ressom}, \bibinfo{person}{Antai Wang}, \bibinfo{person}{Jianhua Xuan},
  \bibinfo{person}{Minetta~C Liu}, \bibinfo{person}{Edmund~A Gehan}, {and}
  \bibinfo{person}{Yue Wang}.} \bibinfo{year}{2008}\natexlab{}.
\newblock \showarticletitle{The properties of high-dimensional data spaces:
  implications for exploring gene and protein expression data}.
\newblock \bibinfo{journal}{\emph{Nature Reviews Cancer}} \bibinfo{volume}{8},
  \bibinfo{number}{1} (\bibinfo{year}{2008}), \bibinfo{pages}{37--49}.
\newblock


\bibitem[\protect\citeauthoryear{Eisner, Rockt{\"{a}}schel, Augenstein,
  Bosnjak, and Riedel}{Eisner et~al\mbox{.}}{2016}]%
        {eisner2016emoji2vec}
\bibfield{author}{\bibinfo{person}{Ben Eisner}, \bibinfo{person}{Tim
  Rockt{\"{a}}schel}, \bibinfo{person}{Isabelle Augenstein},
  \bibinfo{person}{Matko Bosnjak}, {and} \bibinfo{person}{Sebastian Riedel}.}
  \bibinfo{year}{2016}\natexlab{}.
\newblock \showarticletitle{emoji2vec: Learning Emoji Representations from
  their Description}. In \bibinfo{booktitle}{\emph{Proceedings of the
  International Workshop on Natural Language Processing for Social Media
  ({SocialNLP@EMNLP})}}. \bibinfo{publisher}{{ACL}}, \bibinfo{address}{Austin,
  {USA}}, \bibinfo{pages}{48--54}.
\newblock


\bibitem[\protect\citeauthoryear{Engel, Greff, Garth, Bein, Wexler, Hamann, and
  Hagen}{Engel et~al\mbox{.}}{2012}]%
        {engel2012}
\bibfield{author}{\bibinfo{person}{Daniel Engel}, \bibinfo{person}{Klaus
  Greff}, \bibinfo{person}{Christoph Garth}, \bibinfo{person}{Keith Bein},
  \bibinfo{person}{Anthony~S. Wexler}, \bibinfo{person}{Bernd Hamann}, {and}
  \bibinfo{person}{Hans Hagen}.} \bibinfo{year}{2012}\natexlab{}.
\newblock \showarticletitle{Visual Steering and Verification of Mass
  Spectrometry Data Factorization in Air Quality Research}.
\newblock \bibinfo{journal}{\emph{{IEEE} Transactions on Visualization and
  Computer Graphics}} \bibinfo{volume}{18}, \bibinfo{number}{12}
  (\bibinfo{year}{2012}), \bibinfo{pages}{2275--2284}.
\newblock


\bibitem[\protect\citeauthoryear{Engel, Resnick, Roberts, Dieleman, Norouzi,
  Eck, and Simonyan}{Engel et~al\mbox{.}}{2017}]%
        {engel2017neural}
\bibfield{author}{\bibinfo{person}{Jesse~H. Engel}, \bibinfo{person}{Cinjon
  Resnick}, \bibinfo{person}{Adam Roberts}, \bibinfo{person}{Sander Dieleman},
  \bibinfo{person}{Mohammad Norouzi}, \bibinfo{person}{Douglas Eck}, {and}
  \bibinfo{person}{Karen Simonyan}.} \bibinfo{year}{2017}\natexlab{}.
\newblock \showarticletitle{Neural Audio Synthesis of Musical Notes with
  WaveNet Autoencoders}. In \bibinfo{booktitle}{\emph{Proceedings of the
  International Conference on Machine Learning ({ICML})}},
  Vol.~\bibinfo{volume}{70}. \bibinfo{publisher}{{PMLR}},
  \bibinfo{address}{Sydney, Australia}, \bibinfo{pages}{1068--1077}.
\newblock


\bibitem[\protect\citeauthoryear{Ethayarajh, Duvenaud, and Hirst}{Ethayarajh
  et~al\mbox{.}}{2019}]%
        {ethayarajh2019towards}
\bibfield{author}{\bibinfo{person}{Kawin Ethayarajh}, \bibinfo{person}{David
  Duvenaud}, {and} \bibinfo{person}{Graeme Hirst}.}
  \bibinfo{year}{2019}\natexlab{}.
\newblock \showarticletitle{Towards Understanding Linear Word Analogies}. In
  \bibinfo{booktitle}{\emph{Proceedings of the Conference of the Association
  for Computational Linguistics}}. \bibinfo{publisher}{{ACL}},
  \bibinfo{address}{Florence, Italy}, \bibinfo{pages}{3253--3262}.
\newblock


\bibitem[\protect\citeauthoryear{for Disease Control~(CDC) et~al\mbox{.}}{for
  Disease Control~(CDC) et~al\mbox{.}}{1982}]%
        {centers1982update}
\bibfield{author}{\bibinfo{person}{Centers for Disease Control~(CDC)}
  {et~al\mbox{.}}} \bibinfo{year}{1982}\natexlab{}.
\newblock \showarticletitle{Update on acquired immune deficiency syndrome
  (AIDS)--United States.}
\newblock \bibinfo{journal}{\emph{Morbidity and Mortality Weekly Report
  ({MMWR})}} \bibinfo{volume}{31}, \bibinfo{number}{37} (\bibinfo{year}{1982}),
  \bibinfo{pages}{507}.
\newblock


\bibitem[\protect\citeauthoryear{Gleicher, Barve, Yu, and Heimerl}{Gleicher
  et~al\mbox{.}}{2020}]%
        {gleicher2020boxer}
\bibfield{author}{\bibinfo{person}{Michael Gleicher}, \bibinfo{person}{Aditya
  Barve}, \bibinfo{person}{Xinyi Yu}, {and} \bibinfo{person}{Florian Heimerl}.}
  \bibinfo{year}{2020}\natexlab{}.
\newblock \showarticletitle{Boxer: Interactive Comparison of Classifier
  Results}.
\newblock \bibinfo{journal}{\emph{Computer Graphics Forum}}
  \bibinfo{volume}{39}, \bibinfo{number}{3} (\bibinfo{year}{2020}),
  \bibinfo{pages}{181--193}.
\newblock


\bibitem[\protect\citeauthoryear{Ha and Eck}{Ha and Eck}{2018}]%
        {ha2018a}
\bibfield{author}{\bibinfo{person}{David Ha} {and} \bibinfo{person}{Douglas
  Eck}.} \bibinfo{year}{2018}\natexlab{}.
\newblock \showarticletitle{A Neural Representation of Sketch Drawings}. In
  \bibinfo{booktitle}{\emph{Proceedings of the International Conference on
  Learning Representations ({ICLR})}}. \bibinfo{publisher}{OpenReview.net},
  \bibinfo{address}{Vancouver, Canada}.
\newblock


\bibitem[\protect\citeauthoryear{Hamilton, Leskovec, and Jurafsky}{Hamilton
  et~al\mbox{.}}{2016a}]%
        {hamilton2016cultural}
\bibfield{author}{\bibinfo{person}{William~L. Hamilton}, \bibinfo{person}{Jure
  Leskovec}, {and} \bibinfo{person}{Dan Jurafsky}.}
  \bibinfo{year}{2016}\natexlab{a}.
\newblock \showarticletitle{Cultural Shift or Linguistic Drift? Comparing Two
  Computational Measures of Semantic Change}. In
  \bibinfo{booktitle}{\emph{Proceedings of the Conference on Empirical Methods
  in Natural Language Processing ({EMNLP})}}. \bibinfo{publisher}{{ACL}},
  \bibinfo{address}{Austin, {USA}}, \bibinfo{pages}{2116--2121}.
\newblock


\bibitem[\protect\citeauthoryear{Hamilton, Leskovec, and Jurafsky}{Hamilton
  et~al\mbox{.}}{2016b}]%
        {histwords}
\bibfield{author}{\bibinfo{person}{William~L. Hamilton}, \bibinfo{person}{Jure
  Leskovec}, {and} \bibinfo{person}{Dan Jurafsky}.}
  \bibinfo{year}{2016}\natexlab{b}.
\newblock \showarticletitle{Diachronic Word Embeddings Reveal Statistical Laws
  of Semantic Change}. In \bibinfo{booktitle}{\emph{Proceedings of the Annual
  Meeting of the Association for Computational Linguistics ({ACL})}}.
  \bibinfo{publisher}{{ACL}}, \bibinfo{address}{Berlin, Germany}.
\newblock


\bibitem[\protect\citeauthoryear{Heimerl and Gleicher}{Heimerl and
  Gleicher}{2018}]%
        {heimerl2018interactive}
\bibfield{author}{\bibinfo{person}{Florian Heimerl} {and}
  \bibinfo{person}{Michael Gleicher}.} \bibinfo{year}{2018}\natexlab{}.
\newblock \showarticletitle{Interactive Analysis of Word Vector Embeddings}.
\newblock \bibinfo{journal}{\emph{Computer Graphics Forum}}
  \bibinfo{volume}{37}, \bibinfo{number}{3} (\bibinfo{year}{2018}),
  \bibinfo{pages}{253--265}.
\newblock


\bibitem[\protect\citeauthoryear{Heimerl, Kralj, Moller, and Gleicher}{Heimerl
  et~al\mbox{.}}{2020}]%
        {heimerl2020embcomp}
\bibfield{author}{\bibinfo{person}{Florian Heimerl}, \bibinfo{person}{Christoph
  Kralj}, \bibinfo{person}{Torsten Moller}, {and} \bibinfo{person}{Michael
  Gleicher}.} \bibinfo{year}{2020}\natexlab{}.
\newblock \showarticletitle{{embComp}: Visual Interactive Comparison of Vector
  Embeddings}.
\newblock \bibinfo{journal}{\emph{{IEEE} Transactions on Visualization and
  Computer Graphics}} (\bibinfo{year}{2020}), \bibinfo{pages}{1--1}.
\newblock


\bibitem[\protect\citeauthoryear{Hochreiter and Schmidhuber}{Hochreiter and
  Schmidhuber}{1997}]%
        {hochreiter1997long}
\bibfield{author}{\bibinfo{person}{Sepp Hochreiter} {and}
  \bibinfo{person}{J{\"{u}}rgen Schmidhuber}.} \bibinfo{year}{1997}\natexlab{}.
\newblock \showarticletitle{Long Short-Term Memory}.
\newblock \bibinfo{journal}{\emph{Neural Computation}} \bibinfo{volume}{9},
  \bibinfo{number}{8} (\bibinfo{year}{1997}), \bibinfo{pages}{1735--1780}.
\newblock


\bibitem[\protect\citeauthoryear{Hohman, Head, Caruana, DeLine, and
  Drucker}{Hohman et~al\mbox{.}}{2019a}]%
        {gamut}
\bibfield{author}{\bibinfo{person}{Fred Hohman}, \bibinfo{person}{Andrew Head},
  \bibinfo{person}{Rich Caruana}, \bibinfo{person}{Robert DeLine}, {and}
  \bibinfo{person}{Steven~Mark Drucker}.} \bibinfo{year}{2019}\natexlab{a}.
\newblock \showarticletitle{Gamut: {A} Design Probe to Understand How Data
  Scientists Understand Machine Learning Models}. In
  \bibinfo{booktitle}{\emph{Proceedings of the Conference on Human Factors in
  Computing Systems ({CHI})}}. \bibinfo{publisher}{{ACM}},
  \bibinfo{address}{Glasgow, Scotland}, \bibinfo{pages}{579}.
\newblock


\bibitem[\protect\citeauthoryear{Hohman, Kahng, Pienta, and Chau}{Hohman
  et~al\mbox{.}}{2019b}]%
        {hohman2018visual}
\bibfield{author}{\bibinfo{person}{Fred Hohman}, \bibinfo{person}{Minsuk
  Kahng}, \bibinfo{person}{Robert Pienta}, {and} \bibinfo{person}{Duen~Horng
  Chau}.} \bibinfo{year}{2019}\natexlab{b}.
\newblock \showarticletitle{Visual Analytics in Deep Learning: An Interrogative
  Survey for the Next Frontiers}.
\newblock \bibinfo{journal}{\emph{{IEEE} Transactions on Visualization and
  Computer Graphics}} \bibinfo{volume}{25}, \bibinfo{number}{8}
  (\bibinfo{year}{2019}), \bibinfo{pages}{2674--2693}.
\newblock


\bibitem[\protect\citeauthoryear{H{\"{o}}llt, Pezzotti, van Unen, Koning,
  Lelieveldt, and Vilanova}{H{\"{o}}llt et~al\mbox{.}}{2018}]%
        {hollt2017cyteguide}
\bibfield{author}{\bibinfo{person}{Thomas H{\"{o}}llt}, \bibinfo{person}{Nicola
  Pezzotti}, \bibinfo{person}{Vincent van Unen}, \bibinfo{person}{Frits
  Koning}, \bibinfo{person}{Boudewijn P.~F. Lelieveldt}, {and}
  \bibinfo{person}{Anna Vilanova}.} \bibinfo{year}{2018}\natexlab{}.
\newblock \showarticletitle{{CyteGuide}: Visual Guidance for Hierarchical
  Single-Cell Analysis}.
\newblock \bibinfo{journal}{\emph{{IEEE} Transactions on Visualization and
  Computer Graphics}} \bibinfo{volume}{24}, \bibinfo{number}{1}
  (\bibinfo{year}{2018}), \bibinfo{pages}{739--748}.
\newblock


\bibitem[\protect\citeauthoryear{Howard and Ruder}{Howard and Ruder}{2018}]%
        {howard2018universal}
\bibfield{author}{\bibinfo{person}{Jeremy Howard} {and}
  \bibinfo{person}{Sebastian Ruder}.} \bibinfo{year}{2018}\natexlab{}.
\newblock \showarticletitle{Universal Language Model Fine-tuning for Text
  Classification}. In \bibinfo{booktitle}{\emph{Proceedings of the Annual
  Meeting of the Association for Computational Linguistics ({ACL})}}.
  \bibinfo{publisher}{{ACL}}, \bibinfo{address}{Melbourne, Australia},
  \bibinfo{pages}{328--339}.
\newblock


\bibitem[\protect\citeauthoryear{Jin, Barzilay, and Jaakkola}{Jin
  et~al\mbox{.}}{2018}]%
        {jin2018junction}
\bibfield{author}{\bibinfo{person}{Wengong Jin}, \bibinfo{person}{Regina
  Barzilay}, {and} \bibinfo{person}{Tommi~S. Jaakkola}.}
  \bibinfo{year}{2018}\natexlab{}.
\newblock \showarticletitle{Junction Tree Variational Autoencoder for Molecular
  Graph Generation}. In \bibinfo{booktitle}{\emph{Proceedings of the
  International Conference on Machine Learning ({ICML})}},
  Vol.~\bibinfo{volume}{80}. \bibinfo{publisher}{{PMLR}},
  \bibinfo{address}{Stockholm, Sweden}, \bibinfo{pages}{2328--2337}.
\newblock


\bibitem[\protect\citeauthoryear{Jolliffe}{Jolliffe}{1986}]%
        {jolliffe1986principal}
\bibfield{author}{\bibinfo{person}{Ian~T Jolliffe}.}
  \bibinfo{year}{1986}\natexlab{}.
\newblock \showarticletitle{Principal Components in Regression Analysis}.
\newblock In \bibinfo{booktitle}{\emph{Principal Component Analysis}}.
  \bibinfo{publisher}{Springer}, \bibinfo{pages}{129--155}.
\newblock


\bibitem[\protect\citeauthoryear{Kahng, Andrews, Kalro, and Chau}{Kahng
  et~al\mbox{.}}{2018}]%
        {kahng2017cti}
\bibfield{author}{\bibinfo{person}{Minsuk Kahng}, \bibinfo{person}{Pierre~Y.
  Andrews}, \bibinfo{person}{Aditya Kalro}, {and} \bibinfo{person}{Duen
  Horng~(Polo) Chau}.} \bibinfo{year}{2018}\natexlab{}.
\newblock \showarticletitle{ActiVis: Visual Exploration of Industry-Scale Deep
  Neural Network Models}.
\newblock \bibinfo{journal}{\emph{{IEEE} Transactions on Visualization and
  Computer Graphics}} \bibinfo{volume}{24}, \bibinfo{number}{1}
  (\bibinfo{year}{2018}), \bibinfo{pages}{88--97}.
\newblock


\bibitem[\protect\citeauthoryear{Kingma and Ba}{Kingma and Ba}{2015}]%
        {adam}
\bibfield{author}{\bibinfo{person}{Diederik~P. Kingma} {and}
  \bibinfo{person}{Jimmy Ba}.} \bibinfo{year}{2015}\natexlab{}.
\newblock \showarticletitle{Adam: {A} Method for Stochastic Optimization}. In
  \bibinfo{booktitle}{\emph{Proceedings of the International Conference on
  Learning Representations ({ICLR})}}. \bibinfo{address}{San Diego, USA}.
\newblock


\bibitem[\protect\citeauthoryear{Kluyver, Ragan{-}Kelley, P{\'{e}}rez, Granger,
  Bussonnier, Frederic, Kelley, Hamrick, Grout, Corlay, Ivanov, Avila, Abdalla,
  Willing, and Team}{Kluyver et~al\mbox{.}}{2016}]%
        {kluyver2016jupyter}
\bibfield{author}{\bibinfo{person}{Thomas Kluyver}, \bibinfo{person}{Benjamin
  Ragan{-}Kelley}, \bibinfo{person}{Fernando P{\'{e}}rez},
  \bibinfo{person}{Brian~E. Granger}, \bibinfo{person}{Matthias Bussonnier},
  \bibinfo{person}{Jonathan Frederic}, \bibinfo{person}{Kyle Kelley},
  \bibinfo{person}{Jessica~B. Hamrick}, \bibinfo{person}{Jason Grout},
  \bibinfo{person}{Sylvain Corlay}, \bibinfo{person}{Paul Ivanov},
  \bibinfo{person}{Dami{\'{a}}n Avila}, \bibinfo{person}{Safia Abdalla},
  \bibinfo{person}{Carol Willing}, {and} \bibinfo{person}{Jupyter~Development
  Team}.} \bibinfo{year}{2016}\natexlab{}.
\newblock \showarticletitle{Jupyter Notebooks\,---\,a publishing format for
  reproducible computational workflows}. In
  \bibinfo{booktitle}{\emph{Proceedings of the International Conference on
  Electronic Publishing ({ElPub})}}. \bibinfo{publisher}{{IOS} Press},
  \bibinfo{address}{G{\"{o}}ttingen, Germany}, \bibinfo{pages}{87--90}.
\newblock


\bibitem[\protect\citeauthoryear{Koren, Bell, and Volinsky}{Koren
  et~al\mbox{.}}{2009}]%
        {koren2009matrix}
\bibfield{author}{\bibinfo{person}{Yehuda Koren}, \bibinfo{person}{Robert~M.
  Bell}, {and} \bibinfo{person}{Chris Volinsky}.}
  \bibinfo{year}{2009}\natexlab{}.
\newblock \showarticletitle{Matrix Factorization Techniques for Recommender
  Systems}.
\newblock \bibinfo{journal}{\emph{Computer}} \bibinfo{volume}{42},
  \bibinfo{number}{8} (\bibinfo{year}{2009}), \bibinfo{pages}{30--37}.
\newblock


\bibitem[\protect\citeauthoryear{Kusner, Paige, and
  Hern{\'{a}}ndez{-}Lobato}{Kusner et~al\mbox{.}}{2017}]%
        {kusner2017grammar}
\bibfield{author}{\bibinfo{person}{Matt~J. Kusner}, \bibinfo{person}{Brooks
  Paige}, {and} \bibinfo{person}{Jos{\'{e}}~Miguel Hern{\'{a}}ndez{-}Lobato}.}
  \bibinfo{year}{2017}\natexlab{}.
\newblock \showarticletitle{Grammar Variational Autoencoder}. In
  \bibinfo{booktitle}{\emph{Proceedings of the International Conference on
  Machine Learning ({ICML})}}, Vol.~\bibinfo{volume}{70}.
  \bibinfo{publisher}{{PMLR}}, \bibinfo{address}{Sydney, Australia},
  \bibinfo{pages}{1945--1954}.
\newblock


\bibitem[\protect\citeauthoryear{Le and Mikolov}{Le and Mikolov}{2014}]%
        {le2014distributed}
\bibfield{author}{\bibinfo{person}{Quoc~V. Le} {and}
  \bibinfo{person}{Tom{\'{a}}s Mikolov}.} \bibinfo{year}{2014}\natexlab{}.
\newblock \showarticletitle{Distributed Representations of Sentences and
  Documents}. In \bibinfo{booktitle}{\emph{Proceedings of the International
  Conference on Machine Learning ({ICML})}}, Vol.~\bibinfo{volume}{32}.
  \bibinfo{publisher}{JMLR}, \bibinfo{address}{Beijing, China},
  \bibinfo{pages}{1188--1196}.
\newblock


\bibitem[\protect\citeauthoryear{Lekschas, Zhou, Chen, Gehlenborg, Bach, and
  Pfister}{Lekschas et~al\mbox{.}}{2021}]%
        {lekschas2020generic}
\bibfield{author}{\bibinfo{person}{Fritz Lekschas}, \bibinfo{person}{Xinyi
  Zhou}, \bibinfo{person}{Wei Chen}, \bibinfo{person}{Nils Gehlenborg},
  \bibinfo{person}{Benjamin Bach}, {and} \bibinfo{person}{Hanspeter Pfister}.}
  \bibinfo{year}{2021}\natexlab{}.
\newblock \showarticletitle{A Generic Framework and Library for Exploration of
  Small Multiples through Interactive Piling}.
\newblock \bibinfo{journal}{\emph{{IEEE} Transactions on Visualization and
  Computer Graphics}} \bibinfo{volume}{27}, \bibinfo{number}{2}
  (\bibinfo{year}{2021}), \bibinfo{pages}{358--368}.
\newblock


\bibitem[\protect\citeauthoryear{Li, Njotoprawiro, Haleem, Chen, Yi, and Ma}{Li
  et~al\mbox{.}}{2018}]%
        {li2018embeddingvis}
\bibfield{author}{\bibinfo{person}{Quan Li}, \bibinfo{person}{Kristanto~Sean
  Njotoprawiro}, \bibinfo{person}{Hammad Haleem}, \bibinfo{person}{Qiaoan
  Chen}, \bibinfo{person}{Chris Yi}, {and} \bibinfo{person}{Xiaojuan Ma}.}
  \bibinfo{year}{2018}\natexlab{}.
\newblock \showarticletitle{{EmbeddingVis}: {A} Visual Analytics Approach to
  Comparative Network Embedding Inspection}. In
  \bibinfo{booktitle}{\emph{Proceedings of the Conference on Visual Analytics
  Science and Technology ({VAST})}}. \bibinfo{publisher}{{IEEE}},
  \bibinfo{address}{Berlin, Germany}, \bibinfo{pages}{48--59}.
\newblock


\bibitem[\protect\citeauthoryear{Li, Yosinski, Clune, Lipson, and Hopcroft}{Li
  et~al\mbox{.}}{2016}]%
        {li2015convergent}
\bibfield{author}{\bibinfo{person}{Yixuan Li}, \bibinfo{person}{Jason
  Yosinski}, \bibinfo{person}{Jeff Clune}, \bibinfo{person}{Hod Lipson}, {and}
  \bibinfo{person}{John~E. Hopcroft}.} \bibinfo{year}{2016}\natexlab{}.
\newblock \showarticletitle{Convergent Learning: Do different neural networks
  learn the same representations?}. In \bibinfo{booktitle}{\emph{Proceedings of
  the International Conference on Learning Representations ({ICLR})}}.
  \bibinfo{address}{San Juan, Puerto Rico}.
\newblock


\bibitem[\protect\citeauthoryear{Lipton}{Lipton}{2018}]%
        {lipton2018mythos}
\bibfield{author}{\bibinfo{person}{Zachary~C. Lipton}.}
  \bibinfo{year}{2018}\natexlab{}.
\newblock \showarticletitle{The Mythos of Model Interpretability: In Machine
  Learning, the Concept of Interpretability is Both Important and Slippery.}
\newblock \bibinfo{journal}{\emph{Queue}} \bibinfo{volume}{16},
  \bibinfo{number}{3} (\bibinfo{year}{2018}), \bibinfo{pages}{31–57}.
\newblock
\showISSN{1542-7730}


\bibitem[\protect\citeauthoryear{Liu, Bremer, Thiagarajan, Srikumar, Wang,
  Livnat, and Pascucci}{Liu et~al\mbox{.}}{2018}]%
        {liu2017visual}
\bibfield{author}{\bibinfo{person}{Shusen Liu}, \bibinfo{person}{Peer{-}Timo
  Bremer}, \bibinfo{person}{Jayaraman~J. Thiagarajan}, \bibinfo{person}{Vivek
  Srikumar}, \bibinfo{person}{Bei Wang}, \bibinfo{person}{Yarden Livnat}, {and}
  \bibinfo{person}{Valerio Pascucci}.} \bibinfo{year}{2018}\natexlab{}.
\newblock \showarticletitle{Visual Exploration of Semantic Relationships in
  Neural Word Embeddings}.
\newblock \bibinfo{journal}{\emph{{IEEE} Transactions on Visualization and
  Computer Graphics}} \bibinfo{volume}{24}, \bibinfo{number}{1}
  (\bibinfo{year}{2018}), \bibinfo{pages}{553--562}.
\newblock


\bibitem[\protect\citeauthoryear{Liu, Li, Li, Srikumar, Pascucci, and
  Bremer}{Liu et~al\mbox{.}}{2019b}]%
        {liu2018nlize}
\bibfield{author}{\bibinfo{person}{Shusen Liu}, \bibinfo{person}{Zhimin Li},
  \bibinfo{person}{Tao Li}, \bibinfo{person}{Vivek Srikumar},
  \bibinfo{person}{Valerio Pascucci}, {and} \bibinfo{person}{Peer{-}Timo
  Bremer}.} \bibinfo{year}{2019}\natexlab{b}.
\newblock \showarticletitle{{NLIZE:} {A} Perturbation-Driven Visual
  Interrogation Tool for Analyzing and Interpreting Natural Language Inference
  Models}.
\newblock \bibinfo{journal}{\emph{{IEEE} Transactions on Visualization and
  Computer Graphics}} \bibinfo{volume}{25}, \bibinfo{number}{1}
  (\bibinfo{year}{2019}), \bibinfo{pages}{651--660}.
\newblock


\bibitem[\protect\citeauthoryear{Liu, Maljovec, Wang, Bremer, and Pascucci}{Liu
  et~al\mbox{.}}{2017}]%
        {liu2017}
\bibfield{author}{\bibinfo{person}{Shusen Liu}, \bibinfo{person}{Dan Maljovec},
  \bibinfo{person}{Bei Wang}, \bibinfo{person}{Peer{-}Timo Bremer}, {and}
  \bibinfo{person}{Valerio Pascucci}.} \bibinfo{year}{2017}\natexlab{}.
\newblock \showarticletitle{Visualizing High-Dimensional Data: Advances in the
  Past Decade}.
\newblock \bibinfo{journal}{\emph{{IEEE} Transactions on Visualization and
  Computer Graphics}} \bibinfo{volume}{23}, \bibinfo{number}{3}
  (\bibinfo{year}{2017}), \bibinfo{pages}{1249--1268}.
\newblock


\bibitem[\protect\citeauthoryear{Liu, Jun, Li, and Heer}{Liu
  et~al\mbox{.}}{2019a}]%
        {latent-space-cartography}
\bibfield{author}{\bibinfo{person}{Yang Liu}, \bibinfo{person}{Eunice Jun},
  \bibinfo{person}{Qisheng Li}, {and} \bibinfo{person}{Jeffrey Heer}.}
  \bibinfo{year}{2019}\natexlab{a}.
\newblock \showarticletitle{Latent Space Cartography: Visual Analysis of Vector
  Space Embeddings}.
\newblock \bibinfo{journal}{\emph{Computer Graphics Forum}}
  \bibinfo{volume}{38}, \bibinfo{number}{3} (\bibinfo{year}{2019}),
  \bibinfo{pages}{67--78}.
\newblock


\bibitem[\protect\citeauthoryear{Long, Shelhamer, and Darrell}{Long
  et~al\mbox{.}}{2015}]%
        {long2015fully}
\bibfield{author}{\bibinfo{person}{Jonathan Long}, \bibinfo{person}{Evan
  Shelhamer}, {and} \bibinfo{person}{Trevor Darrell}.}
  \bibinfo{year}{2015}\natexlab{}.
\newblock \showarticletitle{Fully Convolutional Networks for Semantic
  Segmentation}. In \bibinfo{booktitle}{\emph{Proceedings of the Conference on
  Computer Vision and Pattern Recognition ({CVPR})}}.
  \bibinfo{publisher}{{IEEE}}, \bibinfo{address}{Boston, {USA}},
  \bibinfo{pages}{3431--3440}.
\newblock


\bibitem[\protect\citeauthoryear{Maas, Daly, Pham, Huang, Ng, and Potts}{Maas
  et~al\mbox{.}}{2011}]%
        {imdbdataset}
\bibfield{author}{\bibinfo{person}{Andrew~L. Maas}, \bibinfo{person}{Raymond~E.
  Daly}, \bibinfo{person}{Peter~T. Pham}, \bibinfo{person}{Dan Huang},
  \bibinfo{person}{Andrew~Y. Ng}, {and} \bibinfo{person}{Christopher Potts}.}
  \bibinfo{year}{2011}\natexlab{}.
\newblock \showarticletitle{Learning Word Vectors for Sentiment Analysis}. In
  \bibinfo{booktitle}{\emph{Proceedings of the Annual Meeting of the
  Association for Computational Linguistics ({ACL})}}.
  \bibinfo{publisher}{{ACL}}, \bibinfo{address}{Portland, {USA}},
  \bibinfo{pages}{142--150}.
\newblock


\bibitem[\protect\citeauthoryear{McInnes, Healy, Saul, and
  Gro{\ss}berger}{McInnes et~al\mbox{.}}{2018}]%
        {mcinnes2018umap}
\bibfield{author}{\bibinfo{person}{Leland McInnes}, \bibinfo{person}{John
  Healy}, \bibinfo{person}{Nathaniel Saul}, {and} \bibinfo{person}{Lukas
  Gro{\ss}berger}.} \bibinfo{year}{2018}\natexlab{}.
\newblock \showarticletitle{{UMAP:} Uniform Manifold Approximation and
  Projection}.
\newblock \bibinfo{journal}{\emph{Journal of Open Source Software}}
  \bibinfo{volume}{3}, \bibinfo{number}{29} (\bibinfo{year}{2018}),
  \bibinfo{pages}{861}.
\newblock


\bibitem[\protect\citeauthoryear{Mikolov, Grave, Bojanowski, Puhrsch, and
  Joulin}{Mikolov et~al\mbox{.}}{2018}]%
        {mikolov2018advances}
\bibfield{author}{\bibinfo{person}{Tom{\'{a}}s Mikolov},
  \bibinfo{person}{Edouard Grave}, \bibinfo{person}{Piotr Bojanowski},
  \bibinfo{person}{Christian Puhrsch}, {and} \bibinfo{person}{Armand Joulin}.}
  \bibinfo{year}{2018}\natexlab{}.
\newblock \showarticletitle{Advances in Pre-Training Distributed Word
  Representations}. In \bibinfo{booktitle}{\emph{Proceedings of the
  International Conference on Language Resources and Evaluation ({LREC})}}.
  \bibinfo{publisher}{{ELRA}}, \bibinfo{address}{Miyazaki, Japan}.
\newblock


\bibitem[\protect\citeauthoryear{Mikolov, Le, and Sutskever}{Mikolov
  et~al\mbox{.}}{2013a}]%
        {mikolov2013exploiting}
\bibfield{author}{\bibinfo{person}{Tom{\'{a}}s Mikolov},
  \bibinfo{person}{Quoc~V. Le}, {and} \bibinfo{person}{Ilya Sutskever}.}
  \bibinfo{year}{2013}\natexlab{a}.
\newblock \bibinfo{title}{Exploiting Similarities among Languages for Machine
  Translation}.
\newblock
\newblock
\showeprint[arxiv]{1309.4168}


\bibitem[\protect\citeauthoryear{Mikolov, Sutskever, Chen, Corrado, and
  Dean}{Mikolov et~al\mbox{.}}{2013b}]%
        {mikolov2013distributed}
\bibfield{author}{\bibinfo{person}{Tom{\'{a}}s Mikolov}, \bibinfo{person}{Ilya
  Sutskever}, \bibinfo{person}{Kai Chen}, \bibinfo{person}{Gregory~S. Corrado},
  {and} \bibinfo{person}{Jeffrey Dean}.} \bibinfo{year}{2013}\natexlab{b}.
\newblock \showarticletitle{Distributed Representations of Words and Phrases
  and their Compositionality}. In \bibinfo{booktitle}{\emph{Advances in Neural
  Information Processing Systems ({NIPS})}}. \bibinfo{publisher}{Curran
  Associates, Inc.}, \bibinfo{address}{Lake Tahoe, {USA}},
  \bibinfo{pages}{3111--3119}.
\newblock


\bibitem[\protect\citeauthoryear{Olah, Satyanarayan, Johnson, Carter, Schubert,
  Ye, and Mordvintsev}{Olah et~al\mbox{.}}{2018}]%
        {olah2018building}
\bibfield{author}{\bibinfo{person}{Chris Olah}, \bibinfo{person}{Arvind
  Satyanarayan}, \bibinfo{person}{Ian Johnson}, \bibinfo{person}{Shan Carter},
  \bibinfo{person}{Ludwig Schubert}, \bibinfo{person}{Katherine Ye}, {and}
  \bibinfo{person}{Alexander Mordvintsev}.} \bibinfo{year}{2018}\natexlab{}.
\newblock \showarticletitle{The Building Blocks of Interpretability}.
\newblock \bibinfo{journal}{\emph{Distill}} (\bibinfo{year}{2018}).
\newblock


\bibitem[\protect\citeauthoryear{Pennington, Socher, and Manning}{Pennington
  et~al\mbox{.}}{2014}]%
        {pennington2014glove}
\bibfield{author}{\bibinfo{person}{Jeffrey Pennington},
  \bibinfo{person}{Richard Socher}, {and} \bibinfo{person}{Christopher~D.
  Manning}.} \bibinfo{year}{2014}\natexlab{}.
\newblock \showarticletitle{{GloVe}: Global Vectors for Word Representation}.
  In \bibinfo{booktitle}{\emph{Proceedings of the Conference on Empirical
  Methods in Natural Language Processing ({EMNLP})}}.
  \bibinfo{publisher}{{ACL}}, \bibinfo{address}{Doha, Qatar},
  \bibinfo{pages}{1532--1543}.
\newblock


\bibitem[\protect\citeauthoryear{Pezzotti, Lelieveldt, van~der Maaten,
  H{\"{o}}llt, Eisemann, and Vilanova}{Pezzotti et~al\mbox{.}}{2017}]%
        {pezzotti2016approximated}
\bibfield{author}{\bibinfo{person}{Nicola Pezzotti}, \bibinfo{person}{Boudewijn
  P.~F. Lelieveldt}, \bibinfo{person}{Laurens van~der Maaten},
  \bibinfo{person}{Thomas H{\"{o}}llt}, \bibinfo{person}{Elmar Eisemann}, {and}
  \bibinfo{person}{Anna Vilanova}.} \bibinfo{year}{2017}\natexlab{}.
\newblock \showarticletitle{Approximated and User Steerable {tSNE} for
  Progressive Visual Analytics}.
\newblock \bibinfo{journal}{\emph{{IEEE} Transactions on Visualization and
  Computer Graphics}} \bibinfo{volume}{23}, \bibinfo{number}{7}
  (\bibinfo{year}{2017}), \bibinfo{pages}{1739--1752}.
\newblock


\bibitem[\protect\citeauthoryear{Rauber, Fadel, Falc{\~{a}}o, and Telea}{Rauber
  et~al\mbox{.}}{2017}]%
        {rauber2016visualizing}
\bibfield{author}{\bibinfo{person}{Paulo~E. Rauber}, \bibinfo{person}{Samuel~G.
  Fadel}, \bibinfo{person}{Alexandre~X. Falc{\~{a}}o}, {and}
  \bibinfo{person}{Alexandru~C. Telea}.} \bibinfo{year}{2017}\natexlab{}.
\newblock \showarticletitle{Visualizing the Hidden Activity of Artificial
  Neural Networks}.
\newblock \bibinfo{journal}{\emph{{IEEE} Transactions on Visualization and
  Computer Graphics}} \bibinfo{volume}{23}, \bibinfo{number}{1}
  (\bibinfo{year}{2017}), \bibinfo{pages}{101--110}.
\newblock


\bibitem[\protect\citeauthoryear{Ribeiro, Singh, and Guestrin}{Ribeiro
  et~al\mbox{.}}{2016}]%
        {ribeiro2016should}
\bibfield{author}{\bibinfo{person}{Marco~Tulio Ribeiro},
  \bibinfo{person}{Sameer Singh}, {and} \bibinfo{person}{Carlos Guestrin}.}
  \bibinfo{year}{2016}\natexlab{}.
\newblock \showarticletitle{"Why Should {I} Trust You?": Explaining the
  Predictions of Any Classifier}. In \bibinfo{booktitle}{\emph{Proceedings of
  the International Conference on Knowledge Discovery and Data Mining
  ({KDD})}}. \bibinfo{publisher}{{ACM}}, \bibinfo{address}{San Francisco, USA},
  \bibinfo{pages}{1135--1144}.
\newblock


\bibitem[\protect\citeauthoryear{Rouditchenko, Boggust, Harwath, Chen, Joshi,
  Thomas, Audhkhasi, Kuehne, Panda, Feris, Kingsbury, Picheny, Torralba, and
  Glass}{Rouditchenko et~al\mbox{.}}{2021}]%
        {rouditchenko2020avlnet}
\bibfield{author}{\bibinfo{person}{Andrew Rouditchenko}, \bibinfo{person}{Angie
  Boggust}, \bibinfo{person}{David Harwath}, \bibinfo{person}{Brian Chen},
  \bibinfo{person}{Dhiraj Joshi}, \bibinfo{person}{Samuel Thomas},
  \bibinfo{person}{Kartik Audhkhasi}, \bibinfo{person}{Hilde Kuehne},
  \bibinfo{person}{Rameswar Panda}, \bibinfo{person}{Rogerio Feris},
  \bibinfo{person}{Brian Kingsbury}, \bibinfo{person}{Michael Picheny},
  \bibinfo{person}{Antonio Torralba}, {and} \bibinfo{person}{James Glass}.}
  \bibinfo{year}{2021}\natexlab{}.
\newblock \showarticletitle{{AVLnet}: Learning Audio-Visual Language
  Representations from Instructional Videos}. In
  \bibinfo{booktitle}{\emph{Proceedings of the Conference of the International
  Speech Communication Association ({INTERSPEECH})}}.
  \bibinfo{publisher}{{ISCA}}, \bibinfo{address}{Brno, Czechia},
  \bibinfo{pages}{1584--1588}.
\newblock


\bibitem[\protect\citeauthoryear{Sacha, Zhang, Sedlmair, Lee, Peltonen,
  Weiskopf, North, and Keim}{Sacha et~al\mbox{.}}{2017}]%
        {sacha2017}
\bibfield{author}{\bibinfo{person}{Dominik Sacha}, \bibinfo{person}{Leishi
  Zhang}, \bibinfo{person}{Michael Sedlmair}, \bibinfo{person}{John~Aldo Lee},
  \bibinfo{person}{Jaakko Peltonen}, \bibinfo{person}{Daniel Weiskopf},
  \bibinfo{person}{Stephen~C. North}, {and} \bibinfo{person}{Daniel~A. Keim}.}
  \bibinfo{year}{2017}\natexlab{}.
\newblock \showarticletitle{Visual Interaction with Dimensionality Reduction:
  {A} Structured Literature Analysis}.
\newblock \bibinfo{journal}{\emph{{IEEE} Transactions on Visualization and
  Computer Graphics}} \bibinfo{volume}{23}, \bibinfo{number}{1}
  (\bibinfo{year}{2017}), \bibinfo{pages}{241--250}.
\newblock


\bibitem[\protect\citeauthoryear{Shiebler}{Shiebler}{2018}]%
        {repcomp}
\bibfield{author}{\bibinfo{person}{Dan Shiebler}.}
  \bibinfo{year}{2018}\natexlab{}.
\newblock \bibinfo{title}{repcomp}.
\newblock
\newblock
\urldef\tempurl%
\url{https://pypi.org/project/repcomp/}
\showURL{%
\tempurl}


\bibitem[\protect\citeauthoryear{Shrikumar, Greenside, and Kundaje}{Shrikumar
  et~al\mbox{.}}{2017}]%
        {shrikumar2017learning}
\bibfield{author}{\bibinfo{person}{Avanti Shrikumar}, \bibinfo{person}{Peyton
  Greenside}, {and} \bibinfo{person}{Anshul Kundaje}.}
  \bibinfo{year}{2017}\natexlab{}.
\newblock \showarticletitle{Learning Important Features Through Propagating
  Activation Differences}. In \bibinfo{booktitle}{\emph{Proceedings of the
  International Conference on Machine Learning ({ICML})}},
  Vol.~\bibinfo{volume}{70}. \bibinfo{publisher}{{PMLR}},
  \bibinfo{address}{Sydney, Australia}, \bibinfo{pages}{3145--3153}.
\newblock


\bibitem[\protect\citeauthoryear{Smilkov, Thorat, Nicholson, Reif, Viégas, and
  Wattenberg}{Smilkov et~al\mbox{.}}{2016}]%
        {smilkov2016embedding}
\bibfield{author}{\bibinfo{person}{Daniel Smilkov}, \bibinfo{person}{Nikhil
  Thorat}, \bibinfo{person}{Charles Nicholson}, \bibinfo{person}{Emily Reif},
  \bibinfo{person}{Fernanda~B. Viégas}, {and} \bibinfo{person}{Martin
  Wattenberg}.} \bibinfo{year}{2016}\natexlab{}.
\newblock \bibinfo{title}{Embedding Projector: Interactive Visualization and
  Interpretation of Embeddings}.
\newblock
\newblock
\showeprint[arxiv]{1611.05469}


\bibitem[\protect\citeauthoryear{Sterling and Irwin}{Sterling and
  Irwin}{2015}]%
        {sterling2015zinc}
\bibfield{author}{\bibinfo{person}{Teague Sterling} {and}
  \bibinfo{person}{John~J. Irwin}.} \bibinfo{year}{2015}\natexlab{}.
\newblock \showarticletitle{{ZINC} 15 -- Ligand Discovery for Everyone}.
\newblock \bibinfo{journal}{\emph{Journal of Chemical Information and
  Modeling}} \bibinfo{volume}{55}, \bibinfo{number}{11} (\bibinfo{year}{2015}),
  \bibinfo{pages}{2324--2337}.
\newblock


\bibitem[\protect\citeauthoryear{Strobelt, Gehrmann, Behrisch, Perer, Pfister,
  and Rush}{Strobelt et~al\mbox{.}}{2019}]%
        {strobelt2019s}
\bibfield{author}{\bibinfo{person}{Hendrik Strobelt},
  \bibinfo{person}{Sebastian Gehrmann}, \bibinfo{person}{Michael Behrisch},
  \bibinfo{person}{Adam Perer}, \bibinfo{person}{Hanspeter Pfister}, {and}
  \bibinfo{person}{Alexander~M. Rush}.} \bibinfo{year}{2019}\natexlab{}.
\newblock \showarticletitle{{Seq2Seq-Vis}: {A} Visual Debugging Tool for
  Sequence-to-Sequence Models}.
\newblock \bibinfo{journal}{\emph{{IEEE} Transactions on Visualization and
  Computer Graphics}} \bibinfo{volume}{25}, \bibinfo{number}{1}
  (\bibinfo{year}{2019}), \bibinfo{pages}{353--363}.
\newblock


\bibitem[\protect\citeauthoryear{Strobelt, Gehrmann, Pfister, and
  Rush}{Strobelt et~al\mbox{.}}{2018}]%
        {strobelt2018lstmvis}
\bibfield{author}{\bibinfo{person}{Hendrik Strobelt},
  \bibinfo{person}{Sebastian Gehrmann}, \bibinfo{person}{Hanspeter Pfister},
  {and} \bibinfo{person}{Alexander~M. Rush}.} \bibinfo{year}{2018}\natexlab{}.
\newblock \showarticletitle{LSTMVis: {A} Tool for Visual Analysis of Hidden
  State Dynamics in Recurrent Neural Networks}.
\newblock \bibinfo{journal}{\emph{{IEEE} Transactions on Visualization and
  Computer Graphics}} \bibinfo{volume}{24}, \bibinfo{number}{1}
  (\bibinfo{year}{2018}), \bibinfo{pages}{667--676}.
\newblock


\bibitem[\protect\citeauthoryear{Sundararajan, Taly, and Yan}{Sundararajan
  et~al\mbox{.}}{2017}]%
        {sundararajan2017axiomatic}
\bibfield{author}{\bibinfo{person}{Mukund Sundararajan}, \bibinfo{person}{Ankur
  Taly}, {and} \bibinfo{person}{Qiqi Yan}.} \bibinfo{year}{2017}\natexlab{}.
\newblock \showarticletitle{Axiomatic Attribution for Deep Networks}. In
  \bibinfo{booktitle}{\emph{Proceedings of the International Conference on
  Machine Learning ({ICML})}}. \bibinfo{publisher}{{PMLR}},
  \bibinfo{address}{Sydney, Australia}, \bibinfo{pages}{3319--3328}.
\newblock


\bibitem[\protect\citeauthoryear{Tan, Zhang, Clarke, and Smucker}{Tan
  et~al\mbox{.}}{2015}]%
        {tan2015lexical}
\bibfield{author}{\bibinfo{person}{Luchen Tan}, \bibinfo{person}{Haotian
  Zhang}, \bibinfo{person}{Charles L.~A. Clarke}, {and}
  \bibinfo{person}{Mark~D. Smucker}.} \bibinfo{year}{2015}\natexlab{}.
\newblock \showarticletitle{Lexical Comparison Between Wikipedia and Twitter
  Corpora by Using Word Embeddings}. In \bibinfo{booktitle}{\emph{Proceedings
  of the Annual Meeting of the Association for Computational Linguistics and
  the International Joint Conference on Natural Language Processing of the
  Asian Federation of Natural Language Processing}}.
  \bibinfo{publisher}{{ACL}}, \bibinfo{address}{Beijing, China},
  \bibinfo{pages}{657--661}.
\newblock


\bibitem[\protect\citeauthoryear{Tufte and Graves-Morris}{Tufte and
  Graves-Morris}{1983}]%
        {tufte1983visual}
\bibfield{author}{\bibinfo{person}{Edward~R Tufte} {and}
  \bibinfo{person}{Peter~R Graves-Morris}.} \bibinfo{year}{1983}\natexlab{}.
\newblock \bibinfo{booktitle}{\emph{The Visual Display of Quantitative
  Information}}. Vol.~\bibinfo{volume}{2}.
\newblock \bibinfo{publisher}{Graphics Press}, \bibinfo{address}{Cheshire,
  {USA}}.
\newblock


\bibitem[\protect\citeauthoryear{Turkay, Kaya, Balcisoy, and Hauser}{Turkay
  et~al\mbox{.}}{2017}]%
        {turkay2016designing}
\bibfield{author}{\bibinfo{person}{Cagatay Turkay}, \bibinfo{person}{Erdem
  Kaya}, \bibinfo{person}{Selim Balcisoy}, {and} \bibinfo{person}{Helwig
  Hauser}.} \bibinfo{year}{2017}\natexlab{}.
\newblock \showarticletitle{Designing Progressive and Interactive Analytics
  Processes for High-Dimensional Data Analysis}.
\newblock \bibinfo{journal}{\emph{{IEEE} Transactions on Visualization and
  Computer Graphics}} \bibinfo{volume}{23}, \bibinfo{number}{1}
  (\bibinfo{year}{2017}), \bibinfo{pages}{131--140}.
\newblock


\bibitem[\protect\citeauthoryear{Turney and Pantel}{Turney and Pantel}{2010}]%
        {turney2010frequency}
\bibfield{author}{\bibinfo{person}{Peter~D. Turney} {and}
  \bibinfo{person}{Patrick Pantel}.} \bibinfo{year}{2010}\natexlab{}.
\newblock \showarticletitle{From Frequency to Meaning: Vector Space Models of
  Semantics}.
\newblock \bibinfo{journal}{\emph{Journal of Artificial Intelligence Research}}
   \bibinfo{volume}{37} (\bibinfo{year}{2010}), \bibinfo{pages}{141--188}.
\newblock


\bibitem[\protect\citeauthoryear{van~der Maaten and Hinton}{van~der Maaten and
  Hinton}{2008}]%
        {maaten2008visualizing}
\bibfield{author}{\bibinfo{person}{Laurens van~der Maaten} {and}
  \bibinfo{person}{Geoffrey Hinton}.} \bibinfo{year}{2008}\natexlab{}.
\newblock \showarticletitle{Visualizing Data using {t-SNE}}.
\newblock \bibinfo{journal}{\emph{Journal of Machine Learning Research}}
  \bibinfo{volume}{9}, \bibinfo{number}{86} (\bibinfo{year}{2008}),
  \bibinfo{pages}{2579--2605}.
\newblock


\bibitem[\protect\citeauthoryear{Wang, Hu, Gu, Hu, Wu, He, and Hopcroft}{Wang
  et~al\mbox{.}}{2018a}]%
        {wang2018towards}
\bibfield{author}{\bibinfo{person}{Liwei Wang}, \bibinfo{person}{Lunjia Hu},
  \bibinfo{person}{Jiayuan Gu}, \bibinfo{person}{Zhiqiang Hu},
  \bibinfo{person}{Yue Wu}, \bibinfo{person}{Kun He}, {and}
  \bibinfo{person}{John~E. Hopcroft}.} \bibinfo{year}{2018}\natexlab{a}.
\newblock \showarticletitle{Towards Understanding Learning Representations: To
  What Extent Do Different Neural Networks Learn the Same Representation}. In
  \bibinfo{booktitle}{\emph{Advances in Neural Information Processing Systems
  ({NeurIPS})}}. \bibinfo{publisher}{Curran Associates, Inc.},
  \bibinfo{address}{Montr{\'{e}}al, Canada}, \bibinfo{pages}{9607--9616}.
\newblock


\bibitem[\protect\citeauthoryear{Wang, Liu, Afzal, Rastegar{-}Mojarad, Wang,
  Shen, Kingsbury, and Liu}{Wang et~al\mbox{.}}{2018b}]%
        {wang2018comparison}
\bibfield{author}{\bibinfo{person}{Yanshan Wang}, \bibinfo{person}{Sijia Liu},
  \bibinfo{person}{Naveed Afzal}, \bibinfo{person}{Majid Rastegar{-}Mojarad},
  \bibinfo{person}{Liwei Wang}, \bibinfo{person}{Feichen Shen},
  \bibinfo{person}{Paul~R. Kingsbury}, {and} \bibinfo{person}{Hongfang Liu}.}
  \bibinfo{year}{2018}\natexlab{b}.
\newblock \showarticletitle{A comparison of word embeddings for the biomedical
  natural language processing}.
\newblock \bibinfo{journal}{\emph{Journal of Biomedical Informatics}}
  \bibinfo{volume}{87} (\bibinfo{year}{2018}), \bibinfo{pages}{12--20}.
\newblock


\bibitem[\protect\citeauthoryear{Wattenberg, Vi{\'e}gas, and
  Johnson}{Wattenberg et~al\mbox{.}}{2016}]%
        {wattenberg2016use}
\bibfield{author}{\bibinfo{person}{Martin Wattenberg},
  \bibinfo{person}{Fernanda Vi{\'e}gas}, {and} \bibinfo{person}{Ian Johnson}.}
  \bibinfo{year}{2016}\natexlab{}.
\newblock \showarticletitle{How to Use {t-SNE} Effectively}.
\newblock \bibinfo{journal}{\emph{Distill}} (\bibinfo{year}{2016}).
\newblock


\bibitem[\protect\citeauthoryear{Weininger}{Weininger}{1988}]%
        {weininger1988smiles}
\bibfield{author}{\bibinfo{person}{David Weininger}.}
  \bibinfo{year}{1988}\natexlab{}.
\newblock \showarticletitle{SMILES, a chemical language and information system.
  1. Introduction to methodology and encoding rules}.
\newblock \bibinfo{journal}{\emph{Journal of Chemical Information and Computer
  Sciences}} \bibinfo{volume}{28}, \bibinfo{number}{1} (\bibinfo{year}{1988}),
  \bibinfo{pages}{31--36}.
\newblock


\bibitem[\protect\citeauthoryear{Willett, Heer, and Agrawala}{Willett
  et~al\mbox{.}}{2007}]%
        {willett2007scented}
\bibfield{author}{\bibinfo{person}{Wesley Willett}, \bibinfo{person}{Jeffrey
  Heer}, {and} \bibinfo{person}{Maneesh Agrawala}.}
  \bibinfo{year}{2007}\natexlab{}.
\newblock \showarticletitle{Scented Widgets: Improving Navigation Cues with
  Embedded Visualizations}.
\newblock \bibinfo{journal}{\emph{{IEEE} Transactions on Visualization and
  Computer Graphics}} \bibinfo{volume}{13}, \bibinfo{number}{6}
  (\bibinfo{year}{2007}), \bibinfo{pages}{1129--1136}.
\newblock


\bibitem[\protect\citeauthoryear{Wongsuphasawat, Moritz, Anand, Mackinlay,
  Howe, and Heer}{Wongsuphasawat et~al\mbox{.}}{2015}]%
        {wongsuphasawat2015voyager}
\bibfield{author}{\bibinfo{person}{Kanit Wongsuphasawat},
  \bibinfo{person}{Dominik Moritz}, \bibinfo{person}{Anushka Anand},
  \bibinfo{person}{Jock~D. Mackinlay}, \bibinfo{person}{Bill Howe}, {and}
  \bibinfo{person}{Jeffrey Heer}.} \bibinfo{year}{2015}\natexlab{}.
\newblock \showarticletitle{Voyager: Exploratory Analysis via Faceted Browsing
  of Visualization Recommendations}.
\newblock \bibinfo{journal}{\emph{{IEEE} Transactions on Visualization and
  Computer Graphics}} \bibinfo{volume}{22}, \bibinfo{number}{1}
  (\bibinfo{year}{2015}), \bibinfo{pages}{649--658}.
\newblock


\bibitem[\protect\citeauthoryear{Wongsuphasawat, Qu, Moritz, Chang, Ouk, Anand,
  Mackinlay, Howe, and Heer}{Wongsuphasawat et~al\mbox{.}}{2017}]%
        {wongsuphasawat2017voyager}
\bibfield{author}{\bibinfo{person}{Kanit Wongsuphasawat},
  \bibinfo{person}{Zening Qu}, \bibinfo{person}{Dominik Moritz},
  \bibinfo{person}{Riley Chang}, \bibinfo{person}{Felix Ouk},
  \bibinfo{person}{Anushka Anand}, \bibinfo{person}{Jock~D. Mackinlay},
  \bibinfo{person}{Bill Howe}, {and} \bibinfo{person}{Jeffrey Heer}.}
  \bibinfo{year}{2017}\natexlab{}.
\newblock \showarticletitle{Voyager 2: Augmenting Visual Analysis with Partial
  View Specifications}. In \bibinfo{booktitle}{\emph{Proceedings of the
  Conference on Human Factors in Computing Systems ({CHI})}}.
  \bibinfo{publisher}{{ACM}}, \bibinfo{address}{Denver, {USA}},
  \bibinfo{pages}{2648--2659}.
\newblock


\bibitem[\protect\citeauthoryear{Xia, Zhang, Song, Chen, Wang, and Liu}{Xia
  et~al\mbox{.}}{2022}]%
        {xia2022}
\bibfield{author}{\bibinfo{person}{Jiazhi Xia}, \bibinfo{person}{Yuchen Zhang},
  \bibinfo{person}{Jie Song}, \bibinfo{person}{Yang Chen},
  \bibinfo{person}{Yunhai Wang}, {and} \bibinfo{person}{Shixia Liu}.}
  \bibinfo{year}{2022}\natexlab{}.
\newblock \showarticletitle{Revisiting Dimensionality Reduction Techniques for
  Visual Cluster Analysis: An Empirical Study}.
\newblock \bibinfo{journal}{\emph{{IEEE} Transactions on Visualization and
  Computer Graphics}} \bibinfo{volume}{28}, \bibinfo{number}{1}
  (\bibinfo{year}{2022}), \bibinfo{pages}{529--539}.
\newblock


\bibitem[\protect\citeauthoryear{Yu, Yang, Bai, Yao, and Rui}{Yu
  et~al\mbox{.}}{2014}]%
        {yu2014visualizing}
\bibfield{author}{\bibinfo{person}{Wei Yu}, \bibinfo{person}{Kuiyuan Yang},
  \bibinfo{person}{Yalong Bai}, \bibinfo{person}{Hongxun Yao}, {and}
  \bibinfo{person}{Yong Rui}.} \bibinfo{year}{2014}\natexlab{}.
\newblock \bibinfo{title}{Visualizing and Comparing Convolutional Neural
  Networks}.
\newblock
\newblock
\showeprint[arxiv]{1412.6631}


\end{thebibliography}
